\documentclass[letterpaper]{sig-alternate-2013}
\usepackage{amsmath,amssymb,amscd,epsfig,amsfonts,rotating}
\usepackage{graphicx}
\usepackage{epstopdf}
\usepackage{subfigure}
\usepackage{multirow}
\usepackage{booktabs}
\usepackage{color}
\usepackage{url}
\usepackage{amsmath,amscd,amsbsy,amssymb,latexsym,url,bm}
\usepackage{epsfig,epstopdf,graphicx,subfigure}
\usepackage{enumitem,balance,mathtools}
\usepackage{wrapfig}
\usepackage{mathrsfs, euscript}
\usepackage{algorithm}
\usepackage{algorithmic}
\usepackage{amssymb}
\usepackage{ifpdf}
\usepackage{alex}

\newfont{\mycrnotice}{ptmr8t at 7pt}
\newfont{\myconfname}{ptmri8t at 7pt}
\let\crnotice\mycrnotice%
\let\confname\myconfname%

\permission{Permission to make digital or hard copies of all or part of this work for personal or classroom use is granted without fee provided that copies are not made or distributed for profit or commercial advantage and that copies bear this notice and the full citation on the first page. Copyrights for components of this work owned by others than ACM must be honored. Abstracting with credit is permitted. To copy otherwise, or republish, to post on servers or to redistribute to lists, requires prior specific permission and/or a fee. Request permissions from Permissions@acm.org.}
\conferenceinfo{KDD'15,}{August 10-13, 2015, Sydney, NSW, Australia.}
\copyrightetc{\copyright~2015 ACM. ISBN \the\acmcopyr}
\crdata{978-1-4503-3664-2/15/08\ ...\$15.00.\\
DOI: http://dx.doi.org/10.1145/2783258.2783269}

\newtheorem{def:def}{Definition}
\newtheorem{thm:thm}{Theorem}
\newtheorem{thm:lm}{Lemma}
\newcommand{\bs}{\boldsymbol}
\newcommand{\tsf}{\textsf}

 \let\oldenumerate\enumerate
 \renewcommand{\enumerate}{
   \oldenumerate
   \setlength{\itemsep}{0pt}
   \setlength{\parskip}{0pt}
   \setlength{\parsep}{0pt}
 }

\title{Statistical Arbitrage Mining for Display Advertising}
\author{
\alignauthor{Weinan Zhang, Jun Wang}\\
	\affaddr{Department of Computer Science, University College London}\\
	\email{ \{w.zhang, j.wang\}@cs.ucl.ac.uk }
}

\begin{document}

\maketitle

\begin{abstract} We study and formulate arbitrage in display advertising. Real-Time Bidding (RTB) mimics stock spot exchanges and utilises computers to algorithmically buy display ads per impression via a real-time auction.  Despite the new automation, the ad markets are still informationally inefficient due to the heavily fragmented marketplaces. Two display impressions with similar or identical effectiveness (e.g., measured by conversion or click-through rates for a targeted audience) may sell for quite different prices at different market segments or pricing schemes. In this paper, we propose a novel data mining paradigm called Statistical Arbitrage Mining (SAM) focusing on mining and exploiting price discrepancies between two pricing schemes. In essence, our SAMer is a meta-bidder that hedges advertisers' risk between CPA (cost per action)-based campaigns and CPM (cost per mille impressions)-based ad inventories; it statistically assesses the potential profit and cost for an incoming CPM bid request against a portfolio of CPA campaigns based on the estimated conversion rate, bid landscape and other statistics learned from historical data. In SAM, (i) functional optimisation is utilised to seek for optimal bidding to maximise the expected arbitrage net profit, and (ii) a portfolio-based risk management solution is leveraged to reallocate bid volume and budget across the set of campaigns to make a risk and return trade-off. We propose to jointly optimise both components in an EM fashion with high efficiency to help the meta-bidder successfully catch the transient statistical arbitrage opportunities in RTB.  Both the offline experiments on a real-world large-scale dataset and online A/B tests on a commercial platform demonstrate the effectiveness of our proposed solution in exploiting arbitrage in various model settings and market environments.
\end{abstract}
\vspace{-10pt}
\vspace{-4pt}

\section{Introduction}\label{sec:intro}

``\emph{Half the money I spend on advertising is wasted; the trouble is I don't know which half.}'' \\
--- John Wanamaker (July 11, 1838 - December 12, 1922)

A popular quotation from John Wanamaker, a pioneer in advertising and department stores, illustrates how difficult it was to quantify the response and performance in advertising a hundred years ago. Over the last twenty years, advancement of the World Wide Web has fundamentally changed this by providing an effective feedback mechanism to measure the response through observing users' search queries, navigation patterns, clicks, conversions etc. Recently,
Real-Time Bidding (RTB) has emerged to be a frontier for Internet advertising \cite{muthukrishnan2009ad,google2011arrival}. It mimics stock spot exchanges and utilises computers to \emph{programmatically} buy display ads in real-time and per impression via an auction mechanism between buyers (advertisers) and sellers (publishers)~\cite{yuan2013real}.

Such automation not only improves efficiency and scales of the buying process across lots of available inventories, but, most importantly, encourages performance driven advertising based on targeted clicks, conversions etc., by using real-time audience data. As a result, ad impressions become more and more \emph{commoditised} in the sense that the effectiveness (quality) of an ad impression does not rely on where it is bought or whom it belongs to any more, but depends on how much it will benefit the campaign target (e.g., underlying Web users' satisfactions and their direct responses)\footnote{Our discussion is limited to performance driven ads and direct responses such as clicks and conversions only, whereas for the purpose of branding, the quality of publishers still play an important role in defining the ad inventory quality.}.

According to the Efficient Market Hypothesis (EMH) in finance, in a perfectly ``efficient'' market, security (such as stock) prices should fully reflect all available information at any time \cite{fama1970efficient}. As such, no \emph{arbitrage} opportunity exists, i.e., one can neither buy securities which are worth more than the selling price, nor sell securities worth less than the selling price without making riskier investment~\cite{hogan2004testing}. However, due to the heavily-fragmented, non-transparent ad marketplaces and the existence of various ad types, e.g., sponsored search, display ads, affiliated networks, and pricing schemes, e.g., cost per mille impressions (CPM), cost per click (CPC), cost per action (CPA), the ad markets are not informationally efficient. In other words, two display opportunities with similar or identical targeted audiences and visit frequency may sell for quite different prices. While exploiting such price discrepancies is still debatable in the advertising field, the following four \emph{arbitrage} situations exist:

\begin{itemize*}
\vspace{-8pt}
\item[I] \textbf{Inter-exchange arbitrage.} Multiple ad exchanges exist. As the supply and demand vary across exchanges for the same user types or targeting rules, there exist intermediary agencies that act as a buyer with low bid in exchange A and as a seller with high reserve price in exchange B in order to make profits \cite{bloomfield2014high}.
\item[II] \textbf{Guaranteed delivery and spot market arbitrage.} Some demand-side platforms (DSPs) offer advertisers the contracts with guaranteed delivery \cite{bharadwaj2010pricing} while buying ad inventories over an RTB exchange with non-guaranteed spot prices \cite{mcafee2012overview}.  Conversely, some ad agencies buy inventories in advance in bulk for fixed ``preferential rates'' from private marketplaces, and then charge a client for their campaigns with the spot prices.
\item[III] \textbf{Publisher volume I/O arbitrage.} A publisher can purchase traffic to her Web page and subsequently make more from ad revenue than the initial inbound click cost. An extreme case is a homepage purely dedicated to host ads: the Million Dollar Homepage\footnote{\url{http://www.milliondollarhomepage.com/}}.
\item[IV] \textbf{Pricing scheme arbitrage.} In RTB, different counter-parties prefer different pricing schemes in order to reduce their risk of deficit \cite{hu2004performance}. CPM is commonly used for RTB auction and preferred by publishers because it is likely to generate stable income from the site volume. By contrast, advertisers focusing on performance are likely to follow CPA and CPC pricing schemes as they are directly related to return on investment (ROI) \cite{edelman2010design}. As such, if the CPM cost to acquire a user conversion is less than the CPA payoff for the conversion, an intermediate agent can earn a positive profit.
\vspace{-5pt}
\end{itemize*}

Scientifically, this is of great interest as it presents a new type of data mining problem, which demands a principled mathematical formulation and novel computational solution to mine and exploit arbitrage opportunities in real-time display advertising.  Commercially and socially, principled ad arbitrage algorithms would not only ensure the business more smooth and risk free (e.g., III \& IV), but also make the ad markets more transparent and informationally more efficient (e.g., I, II \& IV) by connecting otherwise segmented markets to correct the misallocation of risks and prices, and eventually reach an ``arbitrage free'' equilibrium.

In this paper, we formulate Statistical Arbitrage Mining (SAM) and present a solution in the context of display advertising. We focus on modelling discrepancies between CPA-based campaigns and CPM-based ad inventories (IV above), while the arbitrage models for the remaining cases can be obtained analogically.  The studied arbitrage is a stochastic one due to the uncertainty of market supply/demand and users responses.  The probability distribution of the arbitrage net profit from an ad display opportunity is estimated by user response predictors \cite{lee2012estimating} and the bid landscape forecasting models \cite{cui2011bid}, trained on historic large-scale data.  Essentially, the proposed Statistical Arbitrage Miner is a campaign-independent RTB bidder, which assesses the arbitrage opportunity for an incoming CPM bid request against a portfolio of CPA campaigns, then selects a campaign and provides a bid accordingly. Different from previous work on per-campaign RTB bidding strategies \cite{perlich2012bid,zhang2014optimal}, we introduce the concept of \emph{meta-bidder}, which performs the bidding for a portfolio of ad campaigns, similar to a hedge fund holding a set of valid assets in financial markets. In our SAM framework, (i) functional optimisation is utilised to seek for an optimal bidding function to maximise the expected arbitrage net profit, and (ii) a portfolio-based risk management solution is leveraged to reallocate the bidding volume and budget across multiple campaigns to make a trade-off between arbitrage risk and return. We propose to jointly optimise those two components in an EM fashion with high efficiency to make meta-bidder successfully catch the transient statistical arbitrage opportunities in RTB. Experiments on both large-scale datasets and online A/B tests demonstrate the large improvement of our proposed SAM solutions over the state-of-the-art baselines.

\section{Related Work}\label{sec:related-work}

\textbf{Display Advertising Optimisation.}  Before the emerging of the auction-based RTB market, most research work on display advertising optimisation is about ad inventory allocation on behalf of publishers in order to maximise the revenue with the guaranteed delivery constraints \cite{bhalgat2012online,feldman2010online}. The authors in further \cite{bharadwaj2010pricing} propose an automatic model for pricing the guaranteed contracts based on the prices of the targeted individual user visits in a spot market. With the arrival of ad exchange and RTB, a lot of work emerges on auction-based optimisation for display advertising. On the publisher side, the placement-level reserve price optimisation is studied in \cite{yuan2014empirical}. The authors in \cite{ghosh2009bidding} suggest that the publisher could act as a bidder on behalf of its guaranteed contracts so as to make smart inventory allocations among the guaranteed and non-guaranteed contracts. One step further, the pricing model of guaranteed contracts with the alternatives of RTB spot market is proposed in \cite{chen2014dynamic}. On the advertiser side, the bid optimisation for campaign performance improvement is studied. The authors in \cite{lee2013realtime} propose a budget pacing scheme embedded in a campaign conversion revenue optimisation framework to maximise the campaign revenue. The authors in \cite{perlich2012bid,zhang2014optimal} focus on a bidding function formulation to maximise the campaign clicks. Bid landscape forecasting models \cite{cui2011bid} are studied to estimate the campaign's impression volume and cost given a bidding function.

The authors in \cite{cavallo2012display} study auction mechanisms considering arbitrage between CPC and CPM pricing schemes. The study aims for designing an auction mechanism on behalf of the ad exchange and yielding truthful bidding from advertisers and truthful CTR reporting from arbitrageurs. By contrast, our work focuses on developing a statistical method for mining and exploiting arbitrage opportunities between CPA and CPM.

\textbf{Statistical Arbitrage in Finance.}  In financial markets, as a trading strategy, statistical arbitrage is a quantitative approach to security trading. It utilises statistical methods with high-frequency trading systems to detect statistical mispricing of securities caused by market inefficiency to make profit with a large number of transactions \cite{hogan2004testing}.

Drawing an analogy with the statistical arbitrage of security pairs trading \cite{gatev2006pairs} in finance, in our paper, the campaign's CPA contract and its performance in RTB spot markets can be regarded as a pair of correlated securities. Statistically speaking, if the campaign's performance in an RTB market ensures that the average cost to acquire a conversion (i.e., eCPA) is lower than the payoff from the CPA contract, then a statistical arbitrage opportunity exists. Such opportunity could also be considered to be caused by informational inefficiency of the advertising market where the advertisers fail to lower their CPA payoff when their campaigns in RTB spot market have a good performance.

\textbf{Modern Portfolio Theory in Finance.}  As Nobel Prize work \cite{markowitz1952portfolio}, modern portfolio theory (MPT) originates from modelling uncertainty of the return of financial assets. MPT utilises the mean-variance analysis to make an investment solution for any tradeoff between the expected return and the risk, or w.r.t. a reference investment \cite{sharpe1998sharpe}.

Recently, MPT has been introduced into information retrieval (IR) fields to model the expectation and uncertainty of users' preference on the retrieved documents for search engines \cite{wang2009portfolio} or from recommender systems \cite{zhang2013personalize}. To our knowledge, there is no work adopting MPT into the revenue optimisation in online advertising. In our paper, we present a novel way of using MPT and it is naturally integrated into our bid optimisation framework.

\begin{table}[t]
\center
\small
\vspace{-10pt}
\caption{Notations and descriptions.}
\label{tab:notation-des}
\begin{tabular}{rl}
Notation & Description\\
\hline
$\bs{x}$ & The bid request represented by its features.\\
$p_x(\bs{x})$ & The probability density function of $\bs{x}$.\\
$i$ & The $i$th campaign in the meta-bidder portfolio.\\
$M$ & Number of campaigns in the meta-bidder portfolio.\\
$r_i$ & The payoff of campaign $i$ for each conversion.\\
$R$ & The variable of meta-bidder arbitrage net profit.\\
$C$ & The variable of meta-bidder arbitrage cost.\\
$\theta (\bs{x}, i)$ & The predicted CVR if $i$ wins the auction of $\bs{x}$.\\
& We occasionally use $\theta$ to refer to a specific CVR.\\
$p_\theta^i(\theta)$ & The probability density function of CVR $\theta$ for\\
& campaign $i$.\\
$B$ & The meta-bidder total budget.\\
$T$ & The estimated number of bid requests during \\
& the arbitrage period.\\
$b(\theta, r)$ & The bidding function which returns the bid.\\
& $b$ is also used to refer to a specific bid value.\\
$w(b)$ & The probability of winning a bid request with\\
& bid price $b$. \\
$v_i$ & The probability of selecting campaign $i$.\\  
& For multiple campaigns, the campaign selection \\
& probability vector is $\bs{v} = (v_1, v_2, \ldots, v_M)^T$.
\end{tabular}
\end{table}

\section{Statistical Arbitrage Mining}\label{sec:sam}
In this section we formulate and solve the SAM problem in the context of RTB display advertising.  Our paper is intended to be self-contained, but for a detailed introduction of RTB and its ecosystem, we refer to \cite{yuan2013real,zhang2014real}.

\subsection{Problem Definition}
Let us suppose there is an ad agent acting on behalf of advertisers to run their ad campaigns. To hedge advertisers' risk, quite often an ad agent gets paid on the basis of the performance: receive a payoff each time a placed ad eventually leads to a product purchase (cost-per-action, CPA)\footnote{A notable example is \url{mobpartner.com} who explicitly offers payoffs (CPA deals) for anyone who can acquire the needed customers programmatically.}. Note that it remains active research to determine whether and how much a purchase action is attributed to the previously ads shown to the user. In this paper, we adopt the last-touch attribution model commonly used in the industry -- the last ad impression before the user's conversion event is assigned with the full attribution credit \cite{dalessandro2012causally}.
To run the campaigns and place the ads, the agent then goes to the RTB market to purchase ad impressions. In RTB, the ad agent pays the cost for each ad impression displayed (cost-per-mille, CPM) on the basis of second-price auction. In essence, the ad agent is an arbitrageur, making a profit so long as the payoff by conversions (CPA) is higher than the cost (CPM) of acquiring relevant users to making the purchase. Potentially, the agent could in parallel run a large number of campaigns from various advertisers to scale up their profit. Note that the ad agent builds their business by taking the risk from the uncertainty of market competitions and user behaviours. For the entire ad ecosystem, it is healthy as it protects both advertisers and publishers by introducing an intermediary layer that exploits (and ultimately remove) the discrepancies between market segments (in this case, the two pricing schemes, CPA and CPM).

\begin{figure}[t]
  \centering
  \includegraphics[width=\columnwidth]{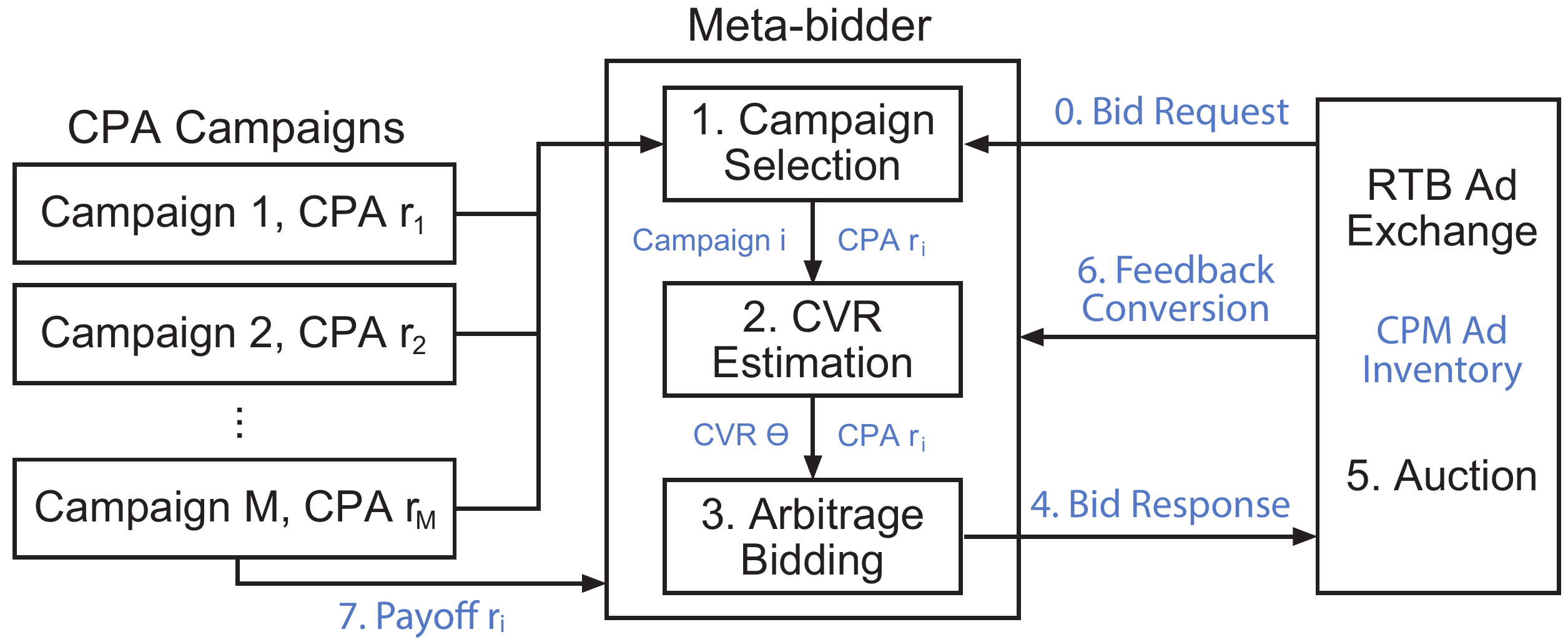}
  \vspace{-15pt}
  \caption{An ad agent running a meta-bidder (arbitrageur) for statistical arbitrage mining.}\label{fig:meta-bidder}
\end{figure}

Traditionally, these arbitrages are accomplished manually. With statistical approaches, it is possible that the above operations can be automatically done by an intelligent meta-bidder across campaigns, where for a certain CPA campaign, the meta-bidder seeks cost-effective ad impressions with high conversion possibility and low market competition.

Mathematically, we formulate the problem below: Suppose there exist $M$ CPA-based campaigns. Each campaign $i$ has set its payoff for a conversion as $r_i$. Over period $T$, the meta-bidder keeps receiving bid requests at time $t\in\{1,\ldots,T\}$, where each bid request is represented with high dimensional feature vector $\bs{x}_t$ and if won, it is charged based on CPM. For each of the incoming bid requests, the Statistical Arbitrage Mining (SAM) problem is to \emph{select a campaign} and \emph{specify its bid} such that over the period $T$ the expected total arbitrage net profit (accumulated payoff minus cost) is maximised.

We consider the following process. When a bid request comes, the meta-bidder samples campaign $i$ with probability $v_i$ to participate the RTB auction, where $\sum_{i=1}^M v_i=1$. Once campaign $i$ is selected, the meta-bidder then estimates its conversion rate (CVR), denoted as $\theta(\bs{x}_t, i)$, i.e., if the ad is placed in this impression, how likely the underlying user will see the ad and eventually convert (purchase)~\cite{lee2012estimating}. After that, the meta-bidder generates the bid price via a bidding function $b(\theta, r_i)$ depending on CVR $\theta(\bs{x}_t, i)$ and conversion payoff $r_i$ \cite{zhang2014optimal}. The notations are summarised in Table~\ref{tab:notation-des};  an illustration on how the SAMer works is given in Figure~\ref{fig:meta-bidder}.

Given campaign selection probability $\bs{v}$ and bidding function $b(\theta, r)$, the meta-bidder's total arbitrage net profit is given by summation over bid requests and campaigns:
\begin{align}                                                                                                                                                                                                   R(\bs{v}, b(\theta, r)) = \sum_{t=1}^{T} \sum_{i=1}^{M} & \Big( \theta(\bs{x}_t, i) r_i - b(\theta(\bs{x}_t, i), r_i) \Big) \cdot \nonumber \\
& w(b(\theta(\bs{x}_t, i), r_i)) v_i, \label{eq:return-variable}
\end{align}
where $w(b)$ is the probability of winning an RTB auction given bid $b$. Product $w(b)v_i$ specifies the probability a campaign is selected \emph{and} wins the auction; $(\theta r_i - b)$ is net profit for the winning campaign. The total cost upper bound is
\begin{align} C(\bs{v}, b(\theta, r)) = \sum_{t=1}^{T} \sum_{i=1}^{M} & b(\theta(\bs{x}_t, i), r_i) w(b(\theta(\bs{x}_t, i), r_i)) v_i,\label{eq:cost-variable}
\end{align}
where bid price $b$ is the maximal possible cost for a campaign to be placed due to the second price auction \cite{yuan2014empirical}.

Next, we need to model how likely we will see an ad impression with feature $\bs{x}_t$ in the future. We assume $\bs{x}_t \sim p_{x}(\bs{x}_t)$; that is for a relatively short period, the bid request feature is drawn from an i.i.d.~built from historic data. The whole model needs to be re-trained periodically with the latest data. Detailed empirical study on the re-training frequency for dynamic arbitrage will be given in Section~\ref{sec:dynamic-artbirage}. Taking the integration over $\bs{x}$ gives the expected net profit:
\begin{align}
  & \mathbb{E}[R(\bs{v}, b(\theta, r))] \nonumber \\
  = & T \int_{\bs{x}} \sum_{i=1}^{M} \Big( \theta(\bs{x}, i) r_i - b(\theta(\bs{x}, i), r_i) \Big) w(b(\theta(\bs{x}, i), r_i)) v_i p_{x}(\bs{x}) d\bs{x} \nonumber \\
= & T \sum_{i=1}^{M} v_i \int_{\theta} \Big( \theta r_i - b(\theta, r_i) \Big) w(b(\theta, r_i)) p_\theta^i(\theta) d\theta,
\end{align}
where $p_\theta^i(\theta(\bs{x}, i)) = p_x(\bs{x})/||\nabla \theta(\bs{x}, i)||$ as there is a deterministic relationship between $\bs{x}$ and its estimated CVR $\theta(\bs{x}, i)$, also given in \cite{zhang2014optimal}. Similarly the total cost is rewritten as
\begin{align}
\mathbb{E}&[C(\bs{v}, b(\theta, r))] = T \sum_{i=1}^{M} v_i \int_{\theta}  b(\theta, r_i) w(b(\theta, r_i)) p_\theta^i(\theta) d\theta.
\end{align}

Finally, the SAM is cast as a constrained optimisation problem: to find campaign selection probability $\bs{v}$ and bidding function $b(\theta, r)$ to maximise the expected arbitrage net profit with budget and risk constraints:
\begin{align}
b_{\text{SAM}}(), \bs{v}^*() = \argmax_{b(),\bs{v}} &~~ \mathbb{E}[R]  \label{eq:optmisation}\\
\text{s.t.} &~~ \mathbb{E}[C] \leq B \label{eq:cost-constraint}\\
&~~ \var[R] \leq h ~~~~~~\label{eq:risk-constraint}\\
&~~ \bs{0} \leq \bs{v} \leq \bs{1} \label{eq:v-pos}\\
&~~ \bs{v}^T \bs{1} = 1, \label{eq:v-sum-1}
\end{align}
where we use variance $\var[R]$ to measure the risk of the net profit and $h$ is a parameter for an upper tolerable risk.

We propose to solve the problem (Eq.~(\ref{eq:optmisation})) in an EM fashion. In particular, the campaign selection probability $\bs{v}$ is regarded as the latent factors to infer and the bidding function $b(\theta, r)$ is regarded as the parameter used to maximise the optimisation target. In E-step, we fix the current estimated bidding function $b(\theta, r)$ and solve the optimal campaign selection probability $\bs{v}$ with the constraints Eqs.~(\ref{eq:risk-constraint}), (\ref{eq:v-pos}), \& (\ref{eq:v-sum-1}). In M-step, we fix the campaign selection probability $\bs{v}$ and seek for the optimal bidding function $b(\theta, r)$ to maximise the target under the budget constraint Eq.~(\ref{eq:cost-constraint}). When the EM iterations get converged, all the constraints are satisfied and the target is maximised. The following Section~\ref{sec:m-step} will describe the detailed solution of optimal bidding function (M-step), and Section~\ref{sec:e-step} will discuss the solution of campaign selection probability $\bs{v}$ (E-step).

\subsection{Optimal Arbitrage Bidding Function}\label{sec:m-step}
With the fixed $\bs{v}$ and the budget constraint at Eq.~(\ref{eq:cost-constraint}), we have a functional optimisation problem in M-step:
\begin{align}
\max_{b()} &~~  T \sum_{i=1}^{M} v_i \int_{\theta}  \Big( \theta r_i - b(\theta, r_i) \Big) w(b(\theta, r_i)) p_\theta^i(\theta) d\theta \label{eq:bid-target} \\
\text{s.t.} &~~ T \sum_{i=1}^{M} v_i \int_{\theta}  b(\theta, r_i) w(b(\theta, r_i)) p_\theta^i(\theta) d\theta \leq B. \label{eq:budget-constraint}
\end{align}

The Lagrangian $\mathcal{L}(b(\theta, r), \lambda)=$
\begin{align}
T \sum_{i=1}^{M} v_i \int_{\theta}  \Big( \theta r_i - (\lambda + 1) b(\theta, r_i) \Big) w(b(\theta, r_i)) p_\theta^i(\theta) d\theta + \lambda B. \label{eq:lagrangian}
\end{align}

Taking its functional derivative w.r.t. $b(\theta, r)$, we have
\begin{align}
\frac{\partial \mathcal{L} (b(\theta, r), \lambda)}{\partial b(\theta, r)} = T \sum_{i=1}^{M} \Big[ & (\theta r_i  - (1 + \lambda) b(\theta, r_i)) \frac{\partial w(b(\theta, r_i))}{\partial b(\theta, r_i)} \nonumber \\
&- (1 + \lambda) w(b(\theta, r_i)) \Big] v_i p_\theta^i(\theta).
\end{align}

A sufficient condition of making this derivative be zero is 
\begin{align}
\Big( \frac{\theta r_i}{1 + \lambda}  -  b(\theta, r_i) \Big) \frac{\partial w(b(\theta, r_i))}{\partial b(\theta, r_i)} =  w(b(\theta, r_i)),\label{eq:optimal-condition}
\end{align}
for all campaign $i$. With the specific functional form of winning function $w(b)$ we can derive the optimal SAM bidding function. Below we show solutions in two special cases.

\begin{figure}[t]
\centering
\subfigure[Winning function]{
\includegraphics[width=0.4\columnwidth]{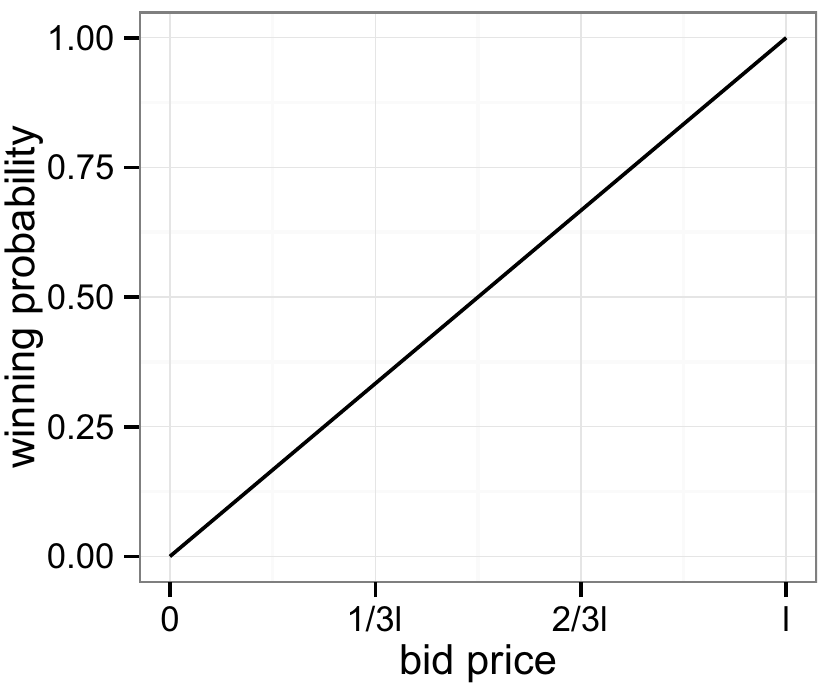}\label{fig:naive-win}}
\subfigure[CVR pdf]{
\includegraphics[width=0.4\columnwidth]{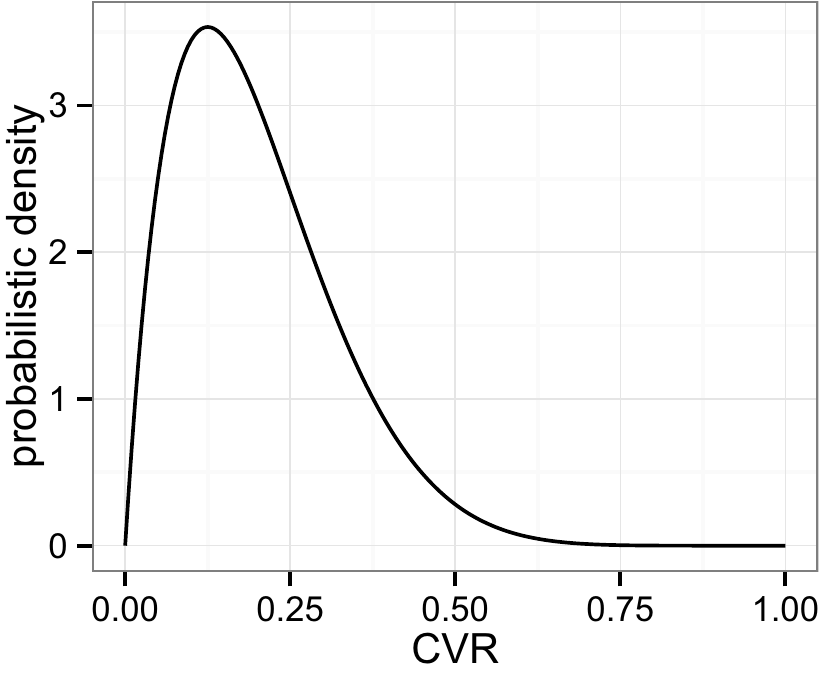}\label{fig:naive-cvr}}
\vspace{-10pt}
\caption{Linear winning function $w(b(\theta))$ and beta CVR pdf $p_\theta(\theta)$.}
\end{figure}

\subsubsection{Uniform Market Price Solution}\label{sec:sam-1}

Here we make a simple example of linear winning function form (see Figure~\ref{fig:naive-win}) based on the assumption of the uniform market price\footnote{Market price refers to the highest bid price amongst the competitors for each auction \cite{amin2012budget}. From a bidder's perspective, it can win an auction if the its bid price is higher than the market price on this auction.} distribution in $[0,l]$:
\begin{align}
w(b(\theta, r)) = \frac{b(\theta, r)}{l}, \label{eq:uniform-win-func}
\end{align}
where the function domain is also $[0,l]$. $l$ is the upper bound of bid price and there is no need to bid higher than $l$.

Replacing Eq.~(\ref{eq:uniform-win-func}) into Eq.~(\ref{eq:optimal-condition}) gives the optimal arbitrage bidding function $b_{\text{sam1}}$ as
\begin{align}
b_{\text{sam1}}(\theta, r) = \frac{r\theta}{2(1+\lambda)}. \label{eq:bid-func-sam1}
\end{align}

To calculate optimal $\lambda$, a sufficient condition of the partial derivative $\partial \mathcal{L}(b(\theta, r),\lambda)/\partial \lambda = 0$ in Eq.~(\ref{eq:lagrangian}) is
\begin{align}
\int_{\theta} b(\theta, r) w(b(\theta, r)) p_\theta(\theta) d\theta &= \frac{B}{T}. \label{eq:general-solve-lambda}
\end{align}

Taking Eqs.~(\ref{eq:uniform-win-func}) and (\ref{eq:bid-func-sam1}) into Eq.~(\ref{eq:general-solve-lambda}) gives
\begin{align}
\frac{r^2}{4(1+\lambda)^2l} \int_{\theta} \theta^2 p_\theta(\theta) d\theta &= \frac{B}{T}, \label{eq:constraint-factorise}
\end{align}
where if we denote $\phi \equiv \int_{\theta} \theta^2 p_\theta(\theta) d\theta$, we have
\begin{align}
\lambda = \frac{r}{2}\sqrt{\frac{T \phi}{B l}} - 1. \label{eq:lambda}
\end{align}

Replacing Eq.~(\ref{eq:lambda}) into Eq.~(\ref{eq:bid-func-sam1}) gives the final solution of bidding function
\begin{align}
b_{\text{sam1}}(\theta, r) = \sqrt{\frac{B l}{T \phi}}\theta, \label{eq:bid-func-sam1-final}
\end{align}
where surprisingly the bidding function does not depend on $r$. This is because the linear forms of $w(b)$ in Eq.~(\ref{eq:uniform-win-func}) and $b_{\text{sam1}}(\theta, r)$ in Eq.~(\ref{eq:bid-func-sam1}) make $\theta$ factorised out from $r/(1+\lambda)$ in Eq.~(\ref{eq:constraint-factorise}), which in turn removes the factor of $r/(1+\lambda)$.
$\phi$ depends on the probabilistic distribution $p_\theta(\theta)$, e.g., the beta distribution \texttt{Beta(2,8)} as shown in Figure~\ref{fig:naive-cvr}, and can be calculated with empirical data.

\subsubsection{Long Tail Market Price Solution}\label{sec:sam-2}

We now consider a more practical winning function used in \cite{zhang2014optimal}, which is based on a long tail market price distribution $p_z(z) = l / (z + l)^2$ with parameter $l$. As such, the winning function is
\begin{align}
w(b(\theta, r)) &= \int_0^{b(\theta, r)} p_z(z)dz =  \frac{b(\theta, r)}{b(\theta, r) + l}. \label{eq:win-func-kdd}
\end{align}

The real-world data analysis on winning bids in \cite{zhang2014optimal} demonstrates the feasibility of adopting the winning function in Eq.~(\ref{eq:win-func-kdd}) in practice.
Taking Eq.~(\ref{eq:win-func-kdd}) into Eq.~(\ref{eq:optimal-condition}) gives the optimal arbitrage bidding function $b_{\text{sam2}}$ as
\begin{align}
b_{\text{sam2}}(\theta, r) = \sqrt{\frac{rl\theta}{1+\lambda} + l^2} - l, \label{eq:bid-func-sam2}
\end{align}
which is in a concave form w.r.t. CVR $\theta$.

\noindent \textbf{Solution of $\lambda$.}
It is possible that the optimal situation does not exhaust the budget and we can leverage the training data to tune the empirically best $\lambda$ as a parameter.
However, if we assume that the bid request volume $T$ is large enough to exhaust the budget, then the optimal case is an equality condition for Eq.~(\ref{eq:budget-constraint}).
To calculate the optimal $\lambda$, the Euler-Lagrange condition of $\lambda$ is Eq.~(\ref{eq:general-solve-lambda}). With Eq.~(\ref{eq:bid-func-sam2}), we explicitly regard $\lambda$ as an input of bidding function $b(\theta, r,\lambda)$ and rewrite $\partial \mathcal{L}(b(\theta, r),\lambda)/\partial \lambda = 0$ from Eq.~(\ref{eq:lagrangian}) as
\begin{align}
\sum_{i=1}^{M} v_i \int_{\theta} b(\theta, r_i, \lambda) w(b(\theta, r_i, \lambda)) p_\theta^i(\theta) d\theta = \frac{B}{T}. \label{eq:solve-lambda}
\end{align}

In most situations except some special cases like Section~\ref{sec:sam-1}, $\lambda$ has no analytic solution. For numeric solution, we can rewrite Eq.~(\ref{eq:solve-lambda}) as
\begin{align}
\sum_{i=1}^{M} v_i \int_{\theta} \Big(b(\theta, r_i, \lambda) w(b(\theta, r_i, \lambda)) - \frac{B}{T}\Big) p_\theta^i(\theta) d\theta = 0,
\end{align}
which has the same solution with the minimisation problem
\begin{align}
\min_{\lambda} \sum_{i=1}^{M} v_i \int_{\theta} \frac{1}{2}\Big(b(\theta, r_i, \lambda) w(b(\theta, r_i, \lambda)) - \frac{B}{T}\Big)^2 p_\theta^i(\theta) d\theta.\nonumber
\end{align}

If we have a very large number $N_i$ of observations of $\theta$'s for each campaign $i$, we can write the above equation as
\begin{align}
\min_{\lambda} \sum_{i=1}^{M} v_i \sum_{k=1}^{N_i} \frac{1}{2}\Big(b(\theta_k^i, r_i, \lambda) w(b(\theta_k^i, r_i, \lambda)) - \frac{B}{T}\Big)^2,\label{eq:batch-min-2}
\end{align}
where we can use (mini-)batch descent or stochastic gradient descent to solve $\lambda$ by the following iteration:
\begin{align}
\lambda \leftarrow& \lambda - \eta \sum_{i=1}^{M} v_i \sum_{k=1}^{N_i} \Big(b(\theta_k^i, r_i, \lambda) w(b(\theta_k^i, r_i, \lambda)) - \frac{B}{T}\Big) \cdot \label{eq:lambda-numeric-solution} \\
&\Big( \frac{\partial b(\theta_k^i, r_i, \lambda)}{\partial \lambda} w(b(\theta_k^i, r_i, \lambda)) + b(\theta_k^i, r_i, \lambda) \frac{\partial w(b(\theta_k^i, r_i, \lambda))}{\partial \lambda} \Big), \nonumber
\end{align}
until convergence. Usually, as $b(\theta, r, \lambda)$ has a monotonic relationship with $\lambda$ and $w(b(\theta, r, \lambda))$ monotonically increases against $b(\theta, r, \lambda)$, $b(\theta_k^i, r, \lambda) w(b(\theta_k^i, r, \lambda))$ has a monotonic relationship with $\lambda$. For example, with the bidding function as Eq.~(\ref{eq:bid-func-sam2}) and the winning function as Eq.~(\ref{eq:win-func-kdd}), the factor $b(\theta_k^i, r, \lambda) w(b(\theta_k^i, r, \lambda))$ decreases monotonically against  $\lambda$, which makes the optimal solution quite easy to find.

\subsection{Optimal Campaign Selection}\label{sec:e-step}
Fixing the resolved optimal arbitrage bidding function $b(\theta, r)$ from previous M-step, we can optimise the campaign selection probability $\bs{v}$ and check whether it is better to reallocate the volume for each campaign.

We here introduce the concept of SAM net profit margin $\gamma$ in RTB display advertising. The net profit margin is the ratio of the net profit of the advertising, either from one campaign or a set of them (meta-bidder), divided by the advertising cost during the corresponding period. In fact, $\gamma = R/C = \text{ROI}-1$.
$\gamma$ is a random variable with expectation and variance. By modelling $\gamma_i$ for each campaign $i$, the optimal campaign selection can be solved by portfolio-based risk management methods.

\subsubsection{Single Campaign}\label{sec:single-campaign}
With optimal arbitrage bidding function $b(\theta, r)$ by Eq.~(\ref{eq:optimal-condition}), we calculate the expectation and variance of the net profit margin $\gamma_i$ for each campaign $i$ by
\begin{align}
\mu_i(b) &= \mathbb{E}[\gamma_i] = \mathbb{E}\Big[ \frac{R_i(\bs{v}_{i=1},b)}{C_i(\bs{v}_{i=1},b)} \Big], \label{eq:single-campaign-mean-general} \\
\sigma_i^2(b) &= \mathbb{E}\Big[\frac{R_i(\bs{v}_{i=1}, b)^2}{C_i(\bs{v}_{i=1}, b)^2} \Big] - \mathbb{E}\Big[\frac{R_i(\bs{v}_{i=1}, b)}{C_i(\bs{v}_{i=1}, b)}\Big]^2,\label{eq:single-campaign-var-general}
\end{align}
where $R_i(\bs{v}_{i=1},b)$ and $C_i(\bs{v}_{i=1},b)$ are as in Eqs.~(\ref{eq:return-variable}) and (\ref{eq:cost-variable}) with $v_i=1$ and $v_j=0$ for all other campaign $j$.
Both $\mu_i(b)$ and $\sigma_i^2(b)$ can be estimated from MCMC methods: (i) repeat $N$ times on sampling $T$ bid requests from the training data and calculate $R_i(\bs{v}_{i=1},b)$ and $C_i(\bs{v}_{i=1},b)$, then (ii) calculate the expectation and variance using these $N$ observations of $R_i(\bs{v}_{i=1},b)$ and $C_i(\bs{v}_{i=1},b)$.

\subsubsection{Campaign Portfolio}
Suppose there are $M$ campaigns in the meta-bidder with CPA contracts.
For each campaign $i$, as discussed in Section~\ref{sec:single-campaign}, there is a variable of achieved net profit margin $\gamma_i$ given the bidding function $b()$, and its expectation is $\mu_i(b)$ and standard deviation is $\sigma_i(b)$. As such, the vector of expected net profit margins for these $M$ campaigns is
\begin{align}
\bs{\mu}(b) = (\mu_1 (b), \mu_2 (b), \ldots, \mu_M (b))^T\label{eq:mu-vector}
\end{align}
and the covariance matrix for the net profit margins of the $M$ campaigns is
$\bs{\Sigma}(b) = \{\sigma_{i,j}(b)\}_{i=1\ldots M,j=1\ldots M}$,
where each element
\begin{align}
\sigma_{i,j}(b) = \beta_{i,j} \sigma_i(b) \sigma_j(b), \label{eq:sigma-element}
\end{align}
where $\beta_{i,j} \in [-1,1]$ is the net profit margin correlation factor between campaign $i$ and $j$, which can be calculated by routine given the net profit margin time series of the two campaigns $i$ and $j$ \cite{markowitz1952portfolio}.

We call such probabilistic campaign combination as \emph{campaign portfolio}.
With the campaign selection probability $\bs{v}$, the campaign portfolio expected net profit margin and its variance are
\begin{align}
\mu_p(\bs{v}, b) = \bs{v}^T \bs{\mu}(b), ~~
\sigma_p^2(\bs{v}, b) = \bs{v}^T \bs{\Sigma}(b) \bs{v}. \label{eq:portfolio-mean-var}
\end{align}

Generally, the arbitrage net profit margin may change w.r.t. the allocated volume: the more bid request volume, the more statistical arbitrage opportunities, and the higher margin. For simplicity, we assume that the net profit margin distribution does not change much w.r.t. the auction volume allocated to the campaign during a short period. The empirical results in Section~\ref{sec:multi-cam-arbitrage} will demonstrate the eligibility of the assumption.


\subsubsection{Campaign Portfolio Optimisation}
The E-step of the original optimisation problem Eq.~(\ref{eq:optmisation}), with the fixed bidding function and constraint Eqs.~(\ref{eq:risk-constraint}), (\ref{eq:v-pos}), \& (\ref{eq:v-sum-1}), can be rewritten by taking the Lagrangian as
\begin{align}
\max_{\bs{v}} ~~& \bs{v}^T \bs{\mu}(b) - \alpha \bs{v}^T \bs{\Sigma}(b) \bs{v}, \label{eq:opt-obj}\\
\text{s.t.} ~~& \bs{v}^T \bs{1} = 1,~~ \bs{0}\leq \bs{v} \leq \bs{1}, \nonumber
\end{align}
where the Lagrangian multiplier $\alpha$ acts as a risk-averse parameter to balance the the expected net profit margin and its variance. This optimisation framework is widely used as portfolio optimisation \cite{wang2009portfolio,zhang2013personalize}.

When the risk, i.e., the variance of the net profit margin, is not considered, $\alpha$ is set as 0. Then the campaign $i$ with the highest $\mu_i(b)$ will be always selected, i.e., $v_i = 1$, while $v_j = 0$ for all other campaigns $j$.

\begin{algorithm}[t]
  \caption{Statistical Arbitrage Mining for Display Ads}\label{algo:sam}
  \small
  \begin{algorithmic}
    \REQUIRE{Meta-bidder winning function $w(b)$}
    \REQUIRE{CTR distribution $p_\theta^i(\theta)$ for each campaign $i$}
    \STATE Initialise $b(\theta, r)= r \theta$ and $\bs{v} = \bs{1}/M$.
    \WHILE {not converged}
     \STATE \textbf{E-step:}
     \STATE ~~Get $\bs{\mu}(b)$ and $\bs{\Sigma}(b)$ by Eq.~(\ref{eq:mu-vector}) and Eq.~(\ref{eq:sigma-element})
     \STATE ~~Solve optimal $\bs{v}$ by Eq.~(\ref{eq:opt-obj})
     \STATE \textbf{M-step:}
     \STATE ~~Get the bidding function form by $w(b)$ and Eq.~(\ref{eq:optimal-condition})
     \STATE ~~Solve $\lambda$ by Eq.~(\ref{eq:lambda-numeric-solution})
     \STATE ~~Update the SAM bidding function $b(\theta, r)$ by Eq.~(\ref{eq:bid-func-sam2})
    \ENDWHILE
    \RETURN $\bs{v}$ and $b(\theta, r)$
  \end{algorithmic}
\end{algorithm}

Finally, the overall operations to get the optimal campaign selection probability $\bs{v}$ and the arbitrage bidding function $b(\theta, r)$ are summarised in Algorithm~\ref{algo:sam}.
Theoretically, just like the EM algorithms for likelihood maximisation, every EM iteration in our case will at least not drop the expected net profit (Eq.~(\ref{eq:optmisation})).
In practice, $\bs{v}$ and $b(\theta, r)$ will get converged within 5 EM iterations. For E-step, the computationally costly parts are the MCMC methods for evaluating the margin of $M$ individual campaign (Eqs.~(\ref{eq:single-campaign-mean-general}) and (\ref{eq:single-campaign-var-general})), where the time complexity is $O(MNT)$, and the campaign correlation calculation ($\beta_{i,j}$ in Eq.~(\ref{eq:sigma-element})), which is $O(M^2NT)$. For M-step, the bidding function is derived with closed form; the calculation of $\lambda$ by numeric descent methods Eq.~(\ref{eq:lambda-numeric-solution}), which depends on the data values but is normally much efficient. The performance in Section~\ref{sec:dynamic-artbirage} will demonstrate the capability of our proposed solution for highly efficient re-training in dynamic arbitrage tasks.

\section{Experiments}\label{sec:exp}
\subsection{Experiment Setting}
\subsubsection{Datasets}\label{sec:dataset}
We conduct our experiments\footnote{The experiment code has been published at \url{https://github.com/wnzhang/rtbarbitrage}.} based on two real-world large-scale bidding logs collected from two DSP companies.

\textbf{iPinYou} RTB dataset was published after iPinYou's global RTB algorithm competition in 2014. This dataset contains the bidding and user feedback log from 9 campaigns during 10 days in 2013, which consists of 64.75M bid records, 19.50M impressions, 14.79K clicks and 16K CNY expense. The dataset disk size is 35GB. According to the data publisher \cite{liao2014ipinyou}, the last three-day data of each campaign is split as the test data and the rest as the training data. More statistics and analysis of the dataset is available in \cite{zhang2014real}.

\textbf{BigTree} RTB dataset is a proprietary dataset from our partner DSP company BigTree Times Co. This dataset is collected from Nov. 2014 to Feb. 2015 for 3 iOS mobile game campaigns. It consists of 10.85M impressions and 46.38K actions\footnote{According to the advertiser's contract, here the action is defined by users' landing on the game's page on app store.} with \$0.083 CPA. We use this dataset to train the model and conduct online A/B test on BigTree DSP during Feb. 2015.

Both datasets are in a record-per-line format, where each line consists of three parts: (i) the features for this auction, e.g., the time, location, IP address, the URL/domain of the publisher, ad slot size, user interest segments etc.; (ii) the auction winning price, which is the threshold of the bid to win this auction; (iii) the user feedback on the ad impression, i.e., click, conversion or not.

\subsubsection{Evaluation Protocol}

\textbf{Evaluation procedure.} We adopt the evaluation procedure similar to the previous work on bid optimisation \cite{zhang2014optimal,zhang2014real}.  In addition, for the evaluation related to the campaign sampling process (via $\bs{v}$), we follow an offline evaluation scheme similar to previous work on evaluating interactive systems \cite{li2011unbiased}. As in the historic data, the user's feedback is only associated with the winning campaign of the auction, there is no corresponding user feedback if a different campaign is sampled. As such, based on the bid request i.i.d. assumption made before, for each round, we first sample a campaign $i$, then pass the next test data record of this campaign to the bid agent for bidding. If there is no more test data left for this campaign, i.e., the bid requests are run out, the test ends.

\textbf{Budget constraints.} It is easy to see that if we set the budget the same as the original total cost in the test log, then simply bidding as much as possible for each auction will exactly run out the budget and get all the logged clicks. In our work, to test the performance against various budget constraints, for each campaign, we respectively run the evaluation test using $1/2, 1/4, 1/8, \ldots, 1/256$ of the original total cost in the test log as the budget.

\textbf{Payoff setting.}
To set up various difficulties in arbitrage, for our offline experiments, we manually adjust the CPA payoff for each iPinYou campaign. Specifically, for each campaign $i$, we set a high and a low CPA payoff in order to test the algorithms' performance under an easy and a hard arbitrage situation, denoted as $r_{i}^{\text{easy}}$ and $r_{i}^{\text{hard}}$, respectively:
\begin{align*}
\begin{array}{c}
r_{i}^{\text{easy}} = \text{eCPA}_i \times 0.8~~\text{and}~~
r_{i}^{\text{hard}} = \text{eCPA}_i \times 0.2,
\end{array}
\end{align*}
where $\text{eCPA}_i$ is the original average cost for acquiring each conversion of campaign $i$ in the training data without any arbitrage strategy. In addition, the conversion data in iPinYou is unavailable for 7 out of 9 campaigns. To have more tests done, we thus regard the user clicks as a proxy for the desired actions (conversions) in our offline experiment.

To complement the offline tests, in our online experiments, we directly adopt the CPA payoff specified by genuine advertisers to test the real business case.

\vspace{5pt}
\subsubsection{Compared Strategies}
We compare the following baseline and state-of-the-art bidding strategies in our experiment. Their parameters are tuned on the training data.
\begin{description*}
\vspace{-5pt}
\item[Constant bidding]
  (\tsf{const}). A constant bid regardless bid requests and campaigns. Though trivial, it is a simple solution widely used by many DSPs.
\item[Random bidding]
    (\tsf{rand}). Randomly choose a bid value in a given range.
\item[Truth-telling bidding]
    (\tsf{truth}). If there is no budget constraint, one should bid the true value for each ad impression, which is CPA$\times$CVR of the impression \cite{lee2012estimating}.
\item[Linear bidding]
    (\tsf{lin}). In \cite{perlich2012bid}, the bid value is linearly proportional to the CVR with the bid scale parameter tuned to maximise the expected conversion number.
\item[Optimal real-time bidding]
    (\tsf{ortb}). This is an optimal bidding strategy proposed in \cite{zhang2014optimal} to maximise clicks. Here we compared the \tsf{ortb1} bid strategy in \cite{zhang2014optimal}.
\item[Statistical arbitrage mining]
    (\tsf{sam1, sam2}). These are the two bidding strategies proposed in this paper: \tsf{sam1} is from Eq.~(\ref{eq:bid-func-sam1}) and \tsf{sam2} is from Eq.~(\ref{eq:bid-func-sam2}), collectively denoted as \tsf{samx}.
\item[SAM with competition modelling]
  (\tsf{sam1c, sam2c}). In a real online environment, the advertisers will tune their bidding strategies according to their campaign performance. If many bidders adopt our \tsf{samx} bidding strategies, it is possible that this may change the market prices. In our offline empirical study, we follow \cite{zhang2012joint} to adopt the \tsf{opt} bidding strategy \cite{chaitanya2012optimal} to simulate the market price changes towards a locally envy-free equilibrium\footnote{The work \cite{chaitanya2012optimal} is on sponsored search with generalised second price auctions. By setting the slot number for each keyword auction as 1 and the CTR as 1.0, the \tsf{opt} bidding strategy can be used for our display advertising scenario.}. Note that this is not for comparing bidding strategies but for comparing auction environment where we would like to check whether our proposed \tsf{samx} algorithms would still make arbitrage profit when the market changes according to our actions. We only compare the performance of \tsf{samx} algorithms with those in the corresponding \tsf{samxc} settings.
\vspace{-5pt}
\end{description*}

For campaign selection strategies, we compare the \tsf{uniform} campaign selection, i.e., $\bs{v} = \bs{1}/M$, and the \tsf{portfolio}-based campaign selection, where \tsf{portfolio} will be denoted as \tsf{greedy} when $\alpha$ in Eq.~(\ref{eq:opt-obj}) is set as 0. The conventional campaign selection scheme based on internal auctions \cite{yuan2013real} will be compared in online A/B test in Section~\ref{sec:online}.

\subsubsection{Evaluation Measures}
We use the net profit as the prime evaluation measure, which is calculated as \texttt{\#conversions * cpa\_payoff - cost}. We also evaluate the net profit margin for each strategy, which is calculated by the net profit divided by the cost. In addition, we report the number of impressions and conversions as well as the cost for each strategy.

\subsection{Single Campaign Arbitrage}\label{sec:single-cam-arbitrage}

\begin{table}[t]
\vspace{-10pt}
\center
\scriptsize
\caption{Single-campaign overall performance.}
\label{tab:stage-1-overall}
\begin{tabular}{rrrrrrr}
\multicolumn{7}{c}{\textbf{Easy payoff, 1/16 budget setting}}\\
bid. & profit & margin & bids & imps. & cnvs. & cost\\
algo. & (CNY) &  & (M) & (K) &  & (CNY)\\ \hline \\[-2ex]
\tsf{const} & 41.77 & 0.21 & 2.68 & 761.91 & 297 & 194.44\\
\tsf{rand} & 19.65 & 0.12 & 2.97 & 612.90 & 223 & 166.60\\
\tsf{truth} & 749.75 & 3.60 & 1.89 & 420.19 & 1,137 & 208.33\\
\tsf{lin} & 845.22 & 3.83 & 2.71 & 531.49 & 1,161 & 220.90\\
\tsf{ortb} & 869.43 & 4.03 & 2.87 & 632.38 & 1,172 & 215.78\\
\tsf{sam1} & 1,141.72 & 6.02 & 3.26 & 471.46 & 1,504 & 189.55\\
\tsf{sam2} & \textbf{1,161.24} & 5.97 & 3.42 & 606.97 & 1,534 & 194.40\\ \hline \\[-2ex]
\tsf{sam1c} & 1,118.61 & 6.10 & 3.24 & 389.09 & 1,473 & 183.34\\
\tsf{sam2c} & 1,141.01 & 5.87 & 3.41 & 563.74 & 1,513 & 194.38\\
\\
\multicolumn{7}{c}{\textbf{Hard payoff, 1/16 budget setting}}\\
bid. & profit & margin & bids & imps. & cnvs. & cost\\
algo. & (CNY) &  & (M) & (K) &  & (CNY)\\ \hline \\[-2ex]
\tsf{const} & -1.40 & -0.25 & 4.10 & 81.55 & 10 & 5.53\\
\tsf{rand} & 1.08 & 10.47 & 4.10 & 8.36 & 4 & 0.10\\
\tsf{truth} & 214.08 & 2.13 & 4.03 & 373.66 & 1,430 & 100.30\\
\tsf{lin} & 45.63 & 0.21 & 2.71 & 531.49 & 1,161 & 220.90\\
\tsf{ortb} & 55.52 & 0.26 & 2.87 & 632.38 & 1,172 & 215.78\\
\tsf{sam1} & 207.34 & 2.29 & 3.89 & 319.77 & 1,328 & 90.59\\
\tsf{sam2} & \textbf{227.76} & 3.77 & 4.10 & 301.99 & 1,326 & 60.47\\ \hline \\[-2ex]
\tsf{sam1c} & 204.73 & 2.25 & 3.88 & 308.44 & 1,322 & 90.98\\
\tsf{sam2c} & 225.95 & 3.70 & 4.10 & 298.51 & 1,322 & 61.13\\
\end{tabular}
\end{table}

\begin{figure}[t]
\centering
\vspace{-14pt}
\subfigure[net profit (easy)]{
\includegraphics[height=1.25in]{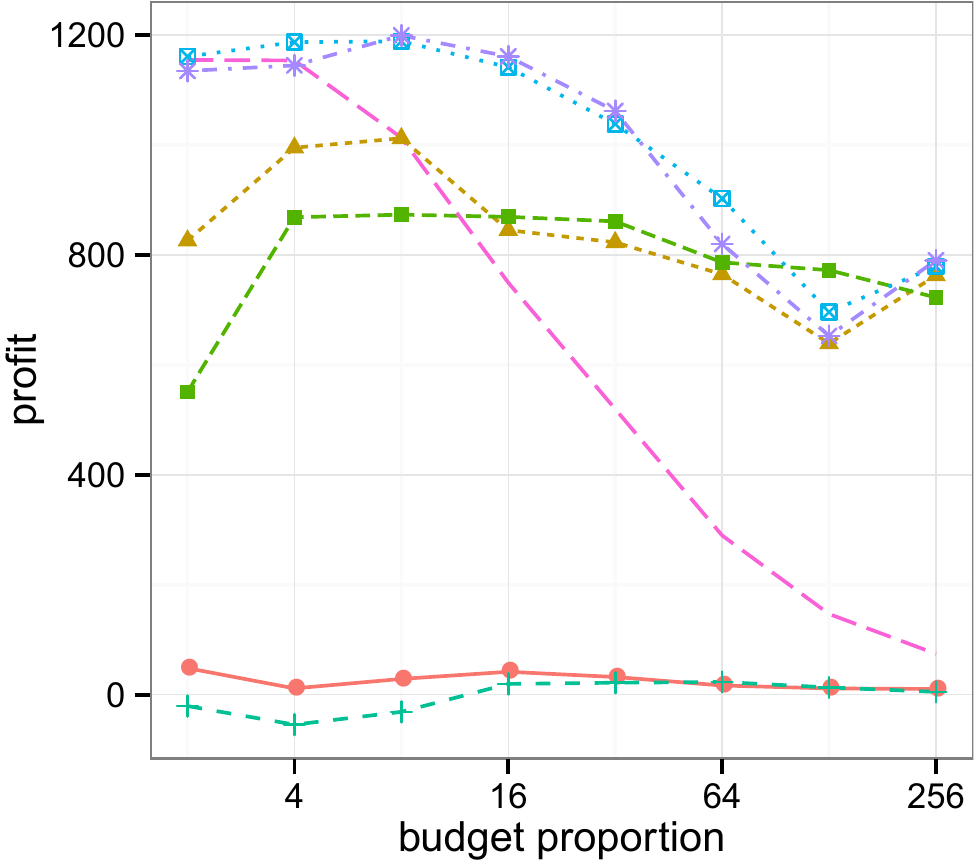}\label{fig:stage-1-profit-easy}}
\subfigure[net profit margin (easy)]{
\includegraphics[height=1.25in]{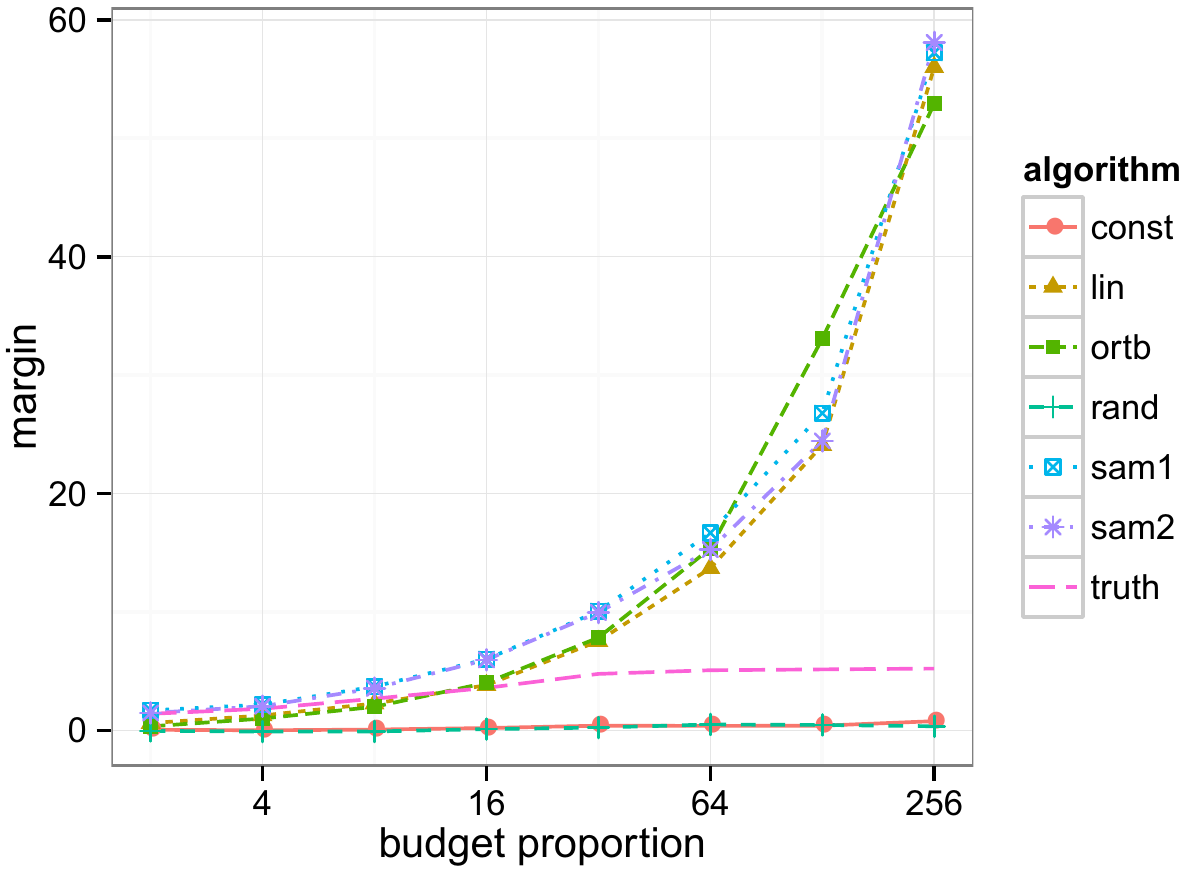}\label{fig:stage-1-margin-easy}}
\vspace{-10pt}
\caption{Single campaign arbitrage performance.}\label{fig:stage-1}
\end{figure}

\begin{figure*}[t]
\centering
\vspace{-7pt}
\subfigure[net profit (easy)]{
\includegraphics[height=1.2in]{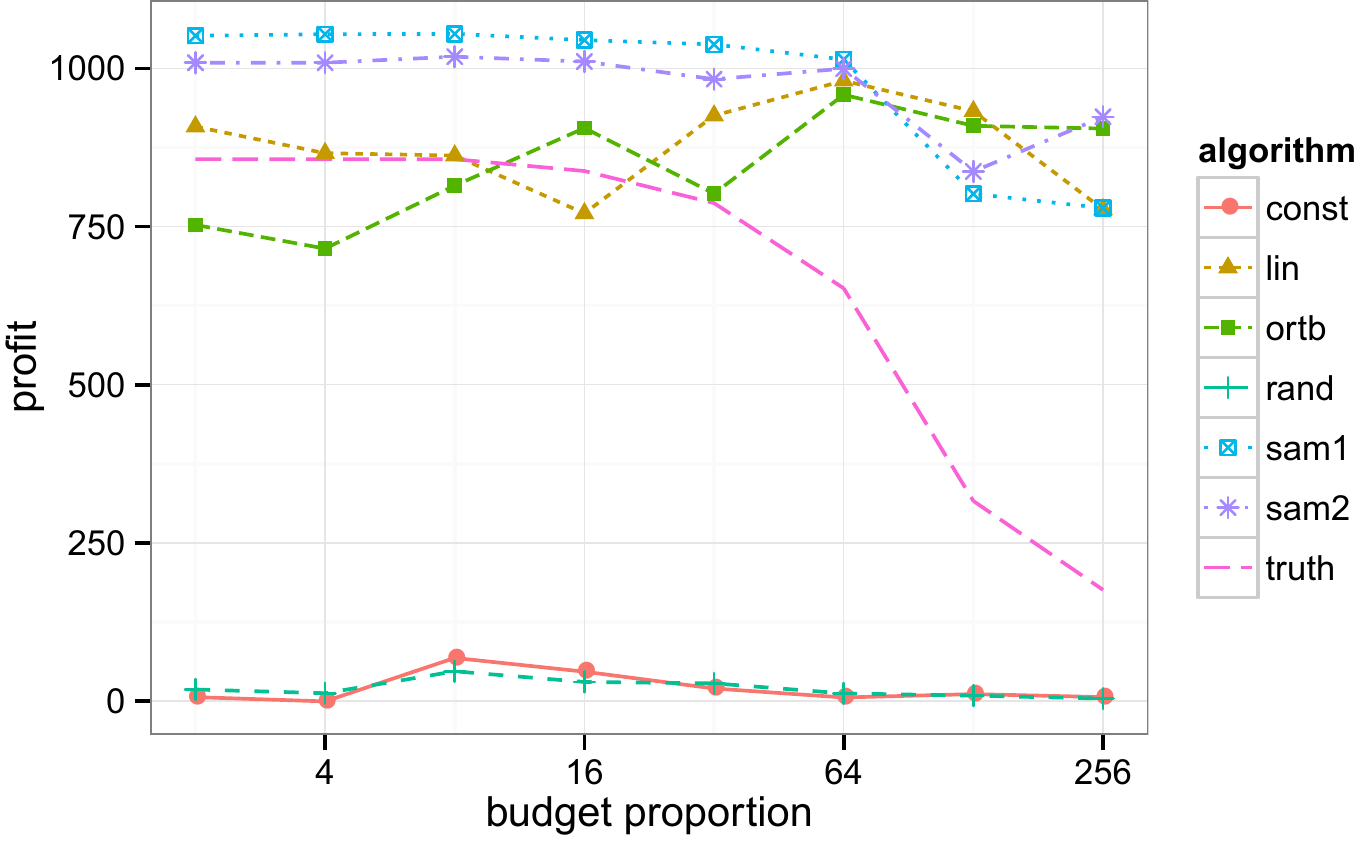}\label{fig:stage-2-profit-easy}}
\subfigure[net profit (hard)]{
\includegraphics[height=1.2in]{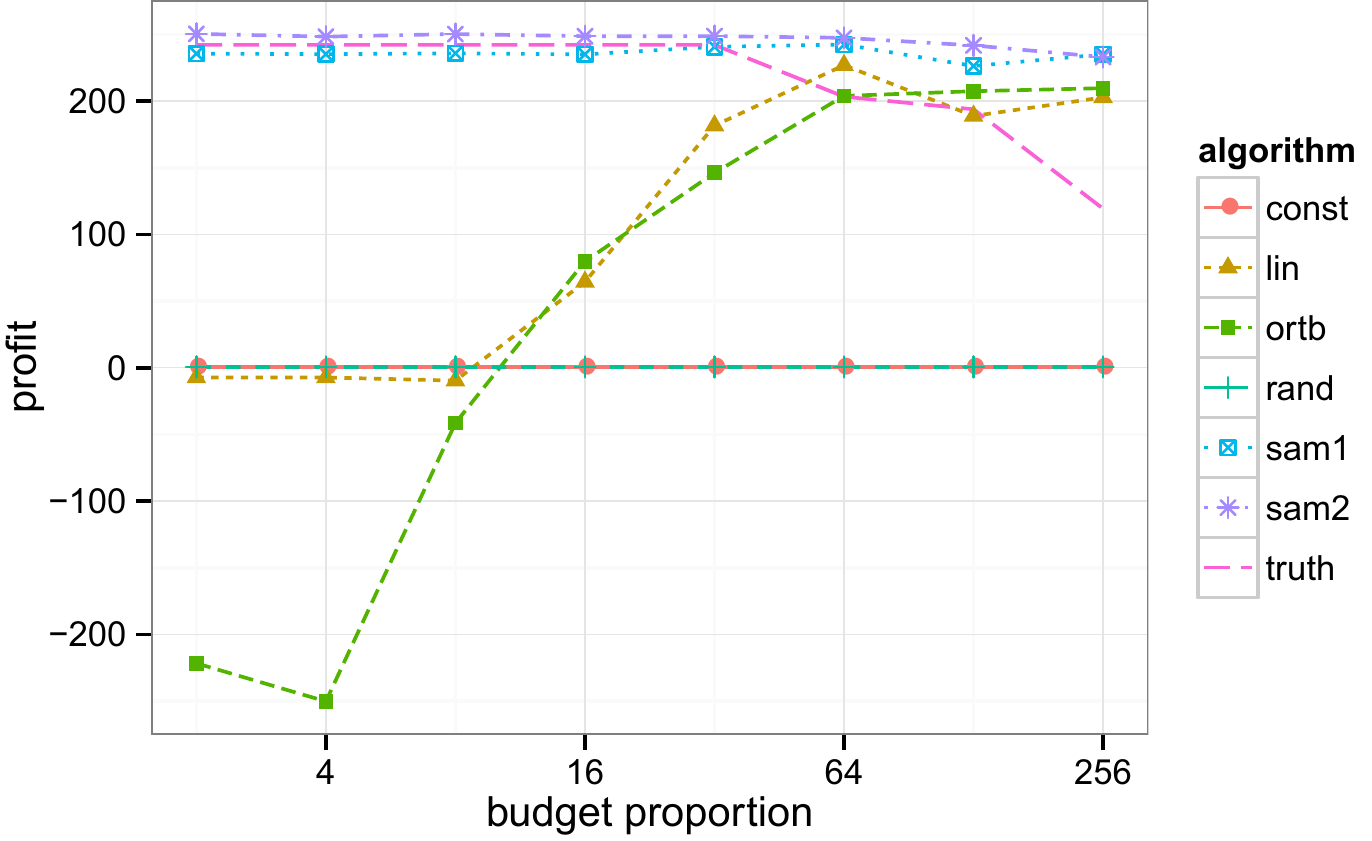}\label{fig:stage-2-profit-hard}}
\subfigure[competition profit drop (easy)]{
\includegraphics[height=1.2in]{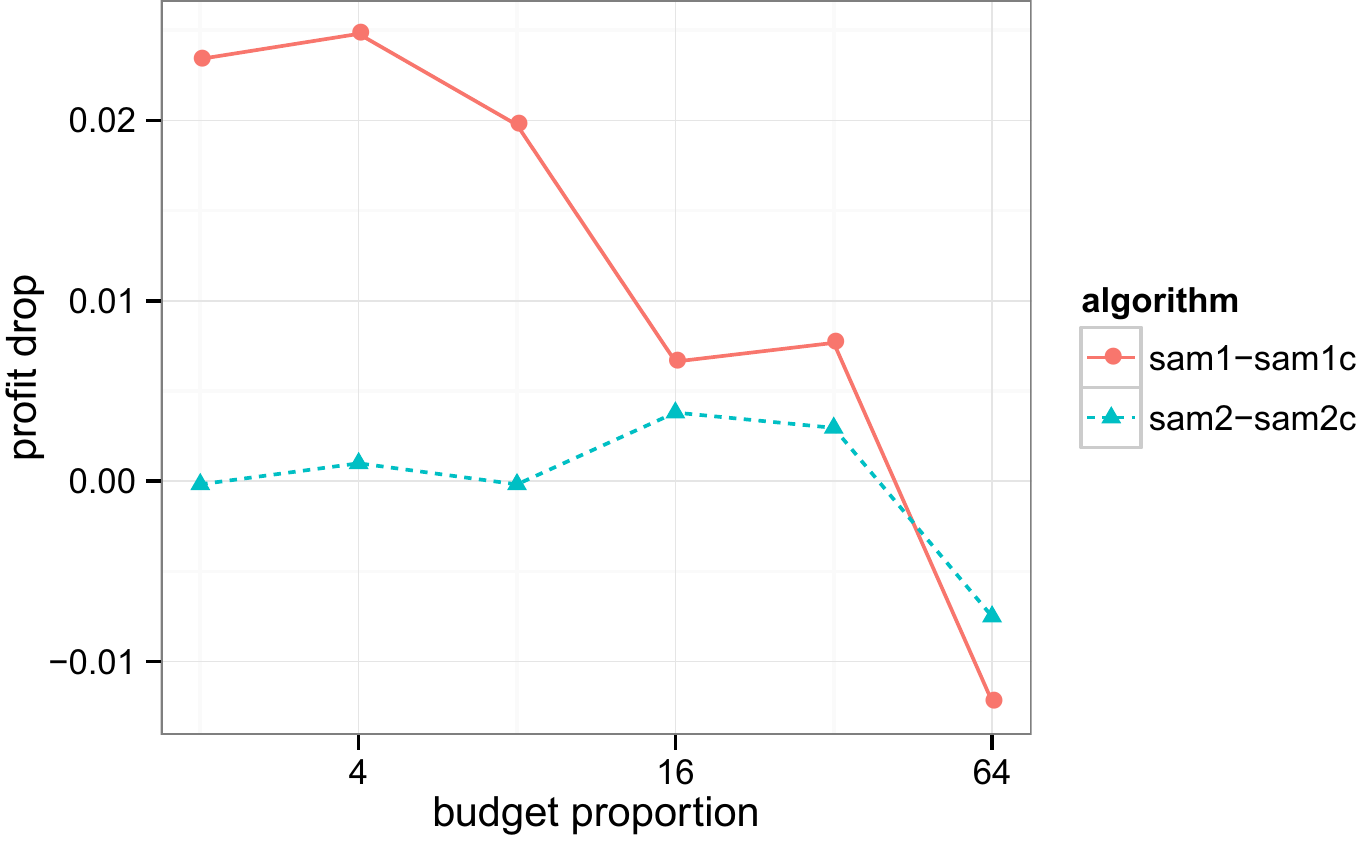}\label{fig:stage-2-profit-drop-easy}}
\subfigure[portfolio $\alpha$ tuning (easy)]{
\includegraphics[height=1.25in]{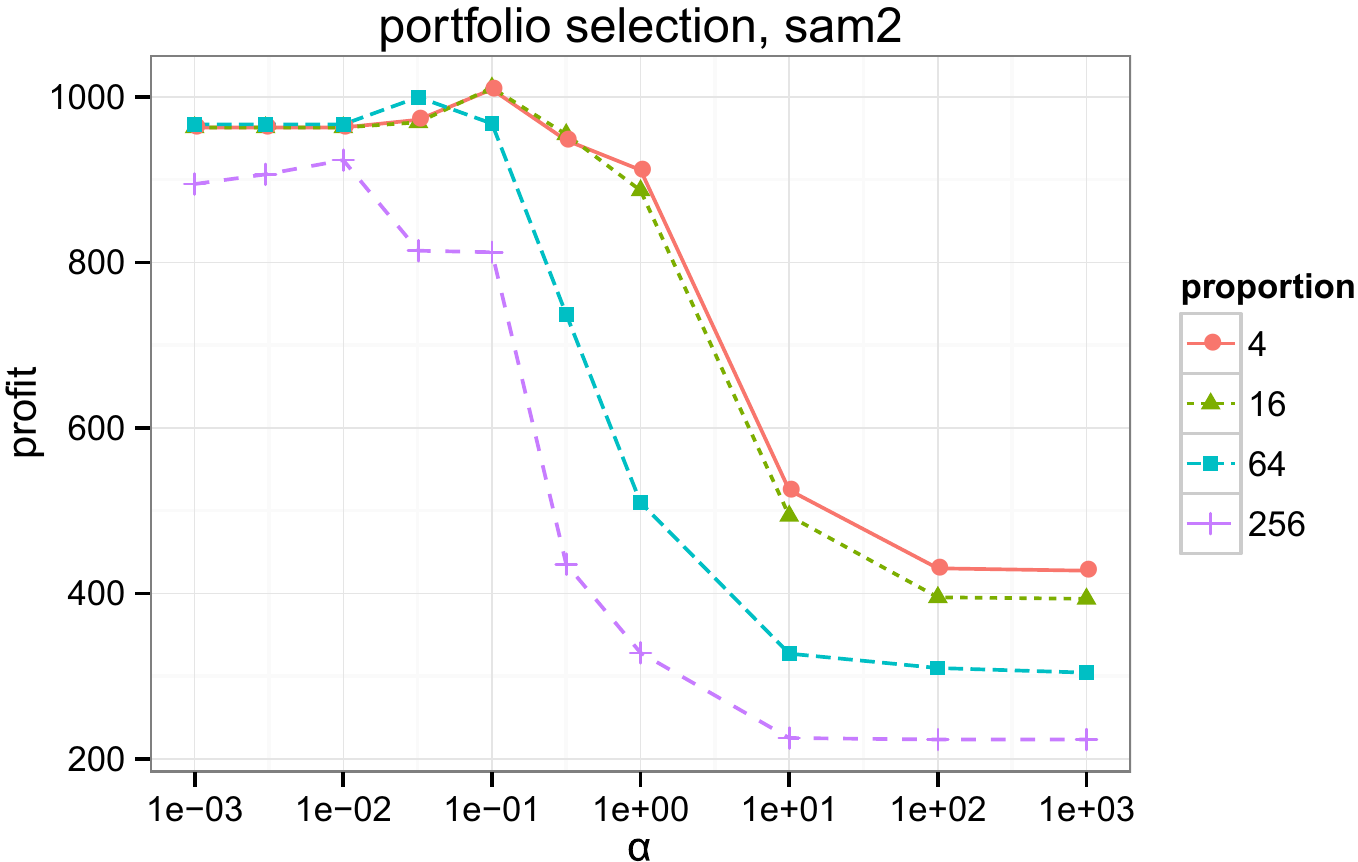}\label{fig:stage-2-alpha-sam2-easy}}
\subfigure[efficient frontier]{
\includegraphics[height=1.25in]{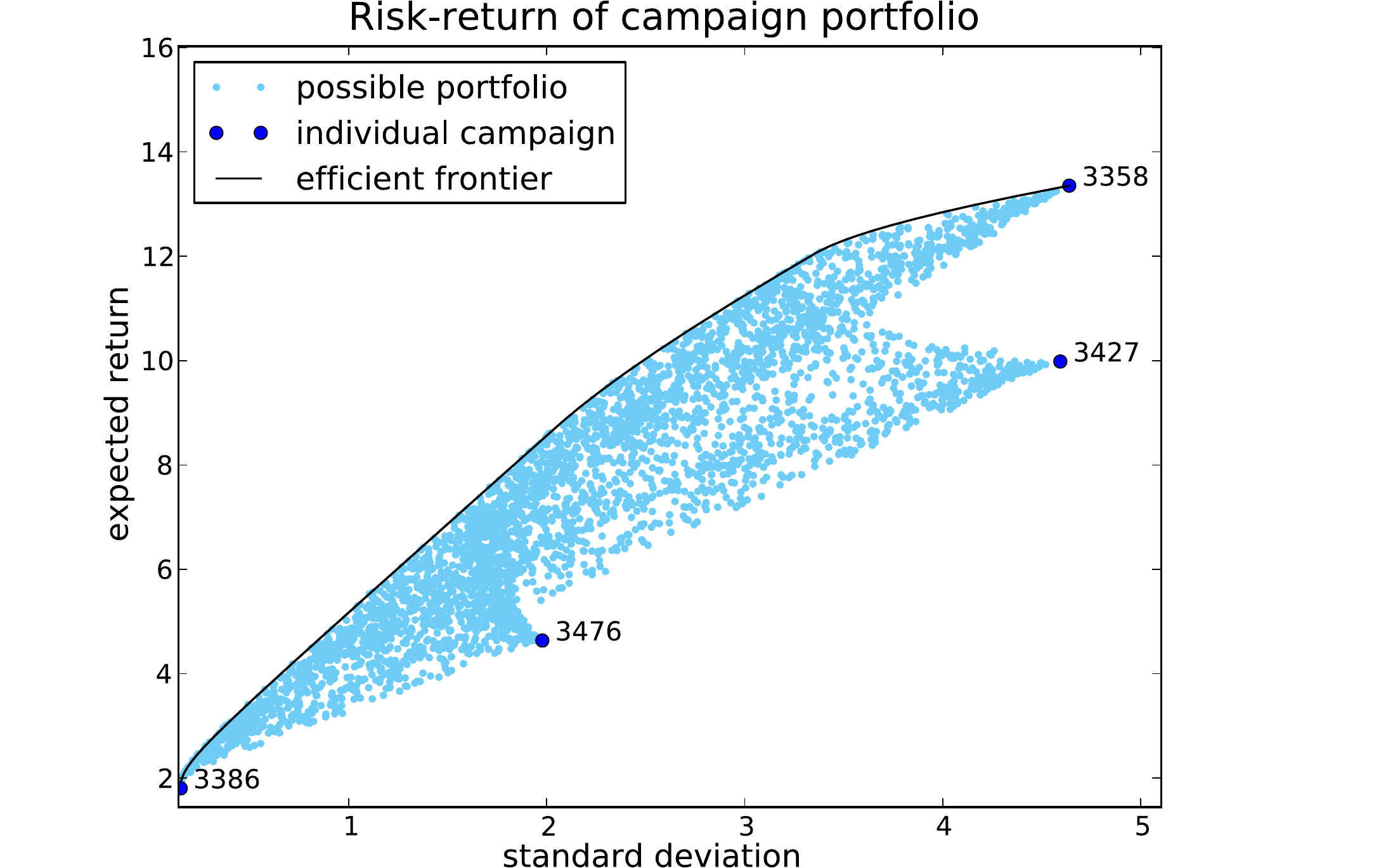}\label{fig:stage-2-efficient-frontier}}
\subfigure[$\bs{v}$ allocation w.r.t. risk]{
\includegraphics[height=1.25in]{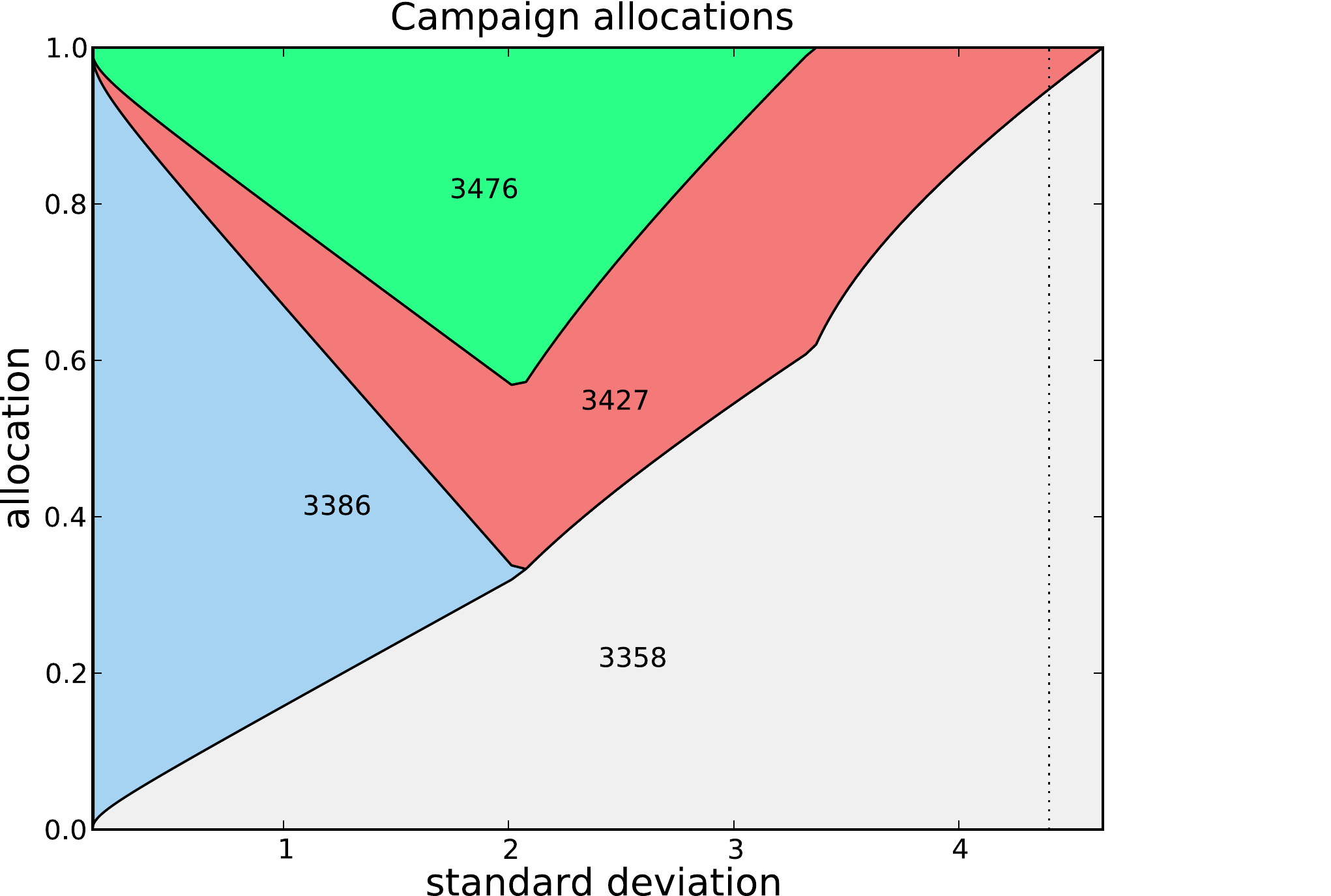}\label{fig:stage-2-allocation}}
\vspace{-10pt}
\caption{Multiple campaign arbitrage performance comparison and campaign portfolio selection analysis.}\label{fig:stage-2}
\vspace{5pt}
\end{figure*}

In Table~\ref{tab:stage-1-overall}, we report the overall performance on the tested 9 campaigns from the iPinYou dataset. We see that \tsf{samx} bidding strategies outperform all others regarding to the net profit. \tsf{sam2} further outperforms \tsf{sam1} particularly in the hard payoff settings due to its more practical winning function. In addition, \tsf{samxc} strategies still make high arbitrage profit with the market competition modelling, which demonstrates the potential of \tsf{samx} strategies in a real market competition environment.

Furthermore, we monitor the performance change on the arbitrage net profit and margin of each algorithm w.r.t. the budget setting in Figure~\ref{fig:stage-1}. For the page limit, we only report the results with the easy payoff setting, while the results on the hard payoff setting are similar. The value on the x-axis means the proportion of the original total cost in the test data divided by the test budget. The higher the proportion is, the less the budget is. From Figure~\ref{fig:stage-1} we have the following observations. (i) \tsf{sam1} and \tsf{sam2} outperform the rest in almost all the profit and margin comparisons with different budget settings.
(ii) Under the higher budget setting, e.g., 2 or 4 budget proportions, \tsf{truth} produces comparable profit as \tsf{samx}. This is because when the budget is abundant, the tight budget constraint (i.e., the equation in Eq.~(\ref{eq:budget-constraint})) is unnecessary to meet in order to maximise the net profit. Under such situation, the bidding problem will get back to the classic second price auction problem, where the truth-telling bidding strategy is optimal \cite{edelman2005internet}. (iii) Under the lower budget setting, e.g., 64, 128 and 256 budget proportions, the profit from \tsf{truth} drops significantly because of the budget constraint is quite important and the optimal bidding strategy is never truth-telling. On the contrary, \tsf{lin} and \tsf{ortb} act almost the same as \tsf{samx}. This is reasonable because under the lower budget settings, the budget is always exhausted. With the cost the same as the budget, the more conversions the more arbitrage profit.

\vspace{-8pt}
\subsection{Multiple Campaign Arbitrage}\label{sec:multi-cam-arbitrage}
We test 6 campaign portfolios from the iPinYou dataset. Each portfolio contains 4 campaigns with the data from the same period. For each portfolio, after the convergence of EM iterations, the empirically optimal $\bs{v}$ and bidding function $b(\theta, r)$ are deployed in the campaign portfolio's test stage, where the auction volume and the budget are set as the same as in the training stage. Compared with the previous single campaign part, this part of experiment focuses more on the campaign portfolio selection, where the \tsf{uniform}, \tsf{greedy} and \tsf{portfolio} selection methods are compared.

\begin{table}[t]
\center
\scriptsize
\vspace{-8pt}
\caption{Multi-campaign overall performance.}
\label{tab:stage-2-overall}
\begin{tabular}{rr|rr|rr}
\multicolumn{2}{c|}{\textbf{strategies}} & \multicolumn{2}{c|}{\textbf{easy payoff}} & \multicolumn{2}{c}{\textbf{hard payoff}}\\
bid. & cam. & profit & margin & profit & margin\\
algo. & select. & (CNY) &  & (CNY) & \\ \hline 
\tsf{lin} & \tsf{greedy} & 501.12 & 6.63 & 68.59 & 0.91\\
\tsf{lin} & \tsf{portfolio} & 925.45 & 13.11 & 181.54 & 2.50\\
\tsf{lin} & \tsf{uniform} & 747.00 & 9.53 & 127.14 & 1.62\\
\tsf{ortb} & \tsf{greedy} & 517.02 & 6.65 & 70.96 & 0.91\\
\tsf{ortb} & \tsf{portfolio} & 802.15 & 10.32 & 146.13 & 1.88\\
\tsf{ortb} & \tsf{uniform} & 765.12 & 9.89 & 133.16 & 1.72\\
\tsf{sam1} & \tsf{greedy} & 966.02 & 20.81 & 230.38 & 11.13\\
\tsf{sam1} & \tsf{portfolio} & \textbf{1,037.98} & 15.84 & 240.63 & 7.96\\
\tsf{sam1} & \tsf{uniform} & 768.38 & 9.78 & 172.43 & 7.57\\
\tsf{sam2} & \tsf{greedy} & 961.68 & 28.73 & 235.31 & 24.00\\
\tsf{sam2} & \tsf{portfolio} & 983.01 & 17.21 & \textbf{248.65} & 13.61\\
\tsf{sam2} & \tsf{uniform} & 774.09 & 10.32 & 168.15 & 5.16\\
\tsf{truth} & \tsf{greedy} & 787.10 & 14.69 & 227.86 & 29.05\\
\tsf{truth} & \tsf{portfolio} & 787.10 & 14.69 & 242.07 & 18.34\\
\tsf{truth} & \tsf{uniform} & 326.57 & 4.14 & 101.12 & 5.36\\
\end{tabular}
\end{table}

The overall results with 1/32 budget setting are reported in Table~\ref{tab:stage-2-overall}. For the comparison among the bidding strategies, \tsf{samx} overall outperforms others in both payoff settings. Figure~\ref{fig:stage-2} provides more detailed analysis. The profit trend against the budget setting, as shown in Figures~\ref{fig:stage-2-profit-easy} and \ref{fig:stage-2-profit-hard}, is consistent with the single campaign setting. The competitor model setting does not significantly drop the arbitrage net profit as shown in Figure~\ref{fig:stage-2-profit-drop-easy}. Specifically, when the budget gets lower, the profit drop percentage gets lower. The reason is that fewer auctions are won with lower budget so that the market does not change much.  To compare campaign selection, Table~\ref{tab:stage-2-overall} shows that \tsf{portfolio} selection constantly outperforms \tsf{uniform} and \tsf{greedy} selection. Compared with \tsf{uniform}, \tsf{greedy} allocates all the auction volume and the budget onto the campaign evaluated as with the highest arbitrage net profit margin, which theoretically maximises the expected net profit. However, the result that \tsf{portfolio} outperforms \tsf{greedy} indicates there exists a return-risk tradeoff point which practically generalises better than the maximum expectation solution. Furthermore, Figure~\ref{fig:stage-2-alpha-sam2-easy} shows the change of total profit from the 6 tested campaign portfolios based on \tsf{sam2} against the portfolio risk-averse parameter $\alpha$ in Eq.~(\ref{eq:opt-obj}). Here setting $\alpha$ as a small enough value is equivalent to the greedy campaign selection. As we can see, as $\alpha$ increases from $10^{-3}$, the net profit first rises to the peak value and then drops significantly. Among the different budget setting, we can observe a trend from Figure~\ref{fig:stage-2-alpha-sam2-easy} that is the more budget, the higher the optimal $\alpha$ is. For 1/256 budget setting, the optimal $\alpha$ is 0.01, while 0.1 is optimal for 1/4 budget setting. This may be due to the fact that more budget brings more auction volume across a longer period, importing more risk, which is required to be carefully hedged.

In addition, we present a case study on a campaign portfolio (3358, 3386, 3427 and 3476 are four campaign IDs). Its return-risk analysis plot is shown in Figure~\ref{fig:stage-2-efficient-frontier} and the corresponding campaign selection probability allocation is shown in Figure~\ref{fig:stage-2-allocation}. In Figure~\ref{fig:stage-2-efficient-frontier} the dark blue points stand for the expected net profit margin and its standard deviation for 4 individual campaigns. As we can see, campaign 3358 has the highest expected margin as well as the highest risk while campaign 3386 is the most stable one but with the lowest expected margin. The best empirical portfolio selection is shown as the vertical dashed line in Figure~\ref{fig:stage-2-allocation}, where 94.9\% auction volume is allocated to campaign 3358 and 4.1\% is allocated to campaign 3427. However, if the meta-bidder is more risk-averse, other two campaigns can be included in order to further reduce the standard deviation.  The parameter $\alpha$ in Eq.~(\ref{eq:opt-obj}) provides a flexible way to adjusting such risk and return trade-off.

\vspace{-3pt}
\subsection{Dynamic Multiple Campaign Arbitrage}\label{sec:dynamic-artbirage}

\begin{figure}[t]
\centering
\vspace{-10pt}
\subfigure[easy payoff setting]{
\includegraphics[height=1.4in]{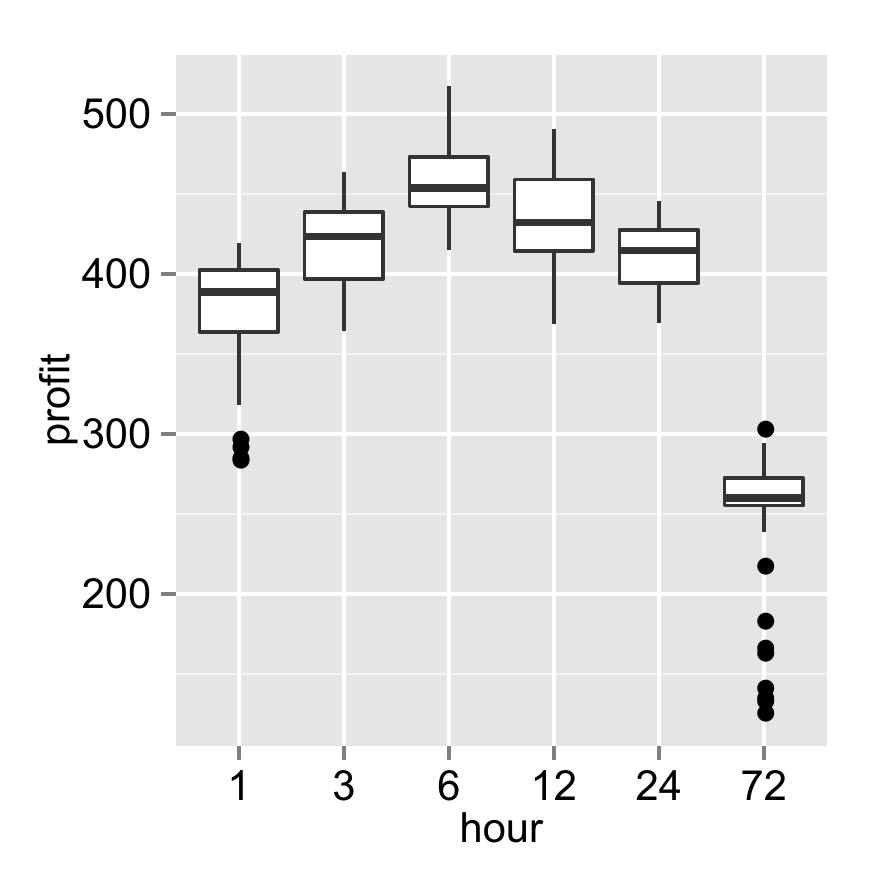}\label{fig:stage-3-profit-hour-easy}}
\subfigure[hard payoff setting]{
\includegraphics[height=1.4in]{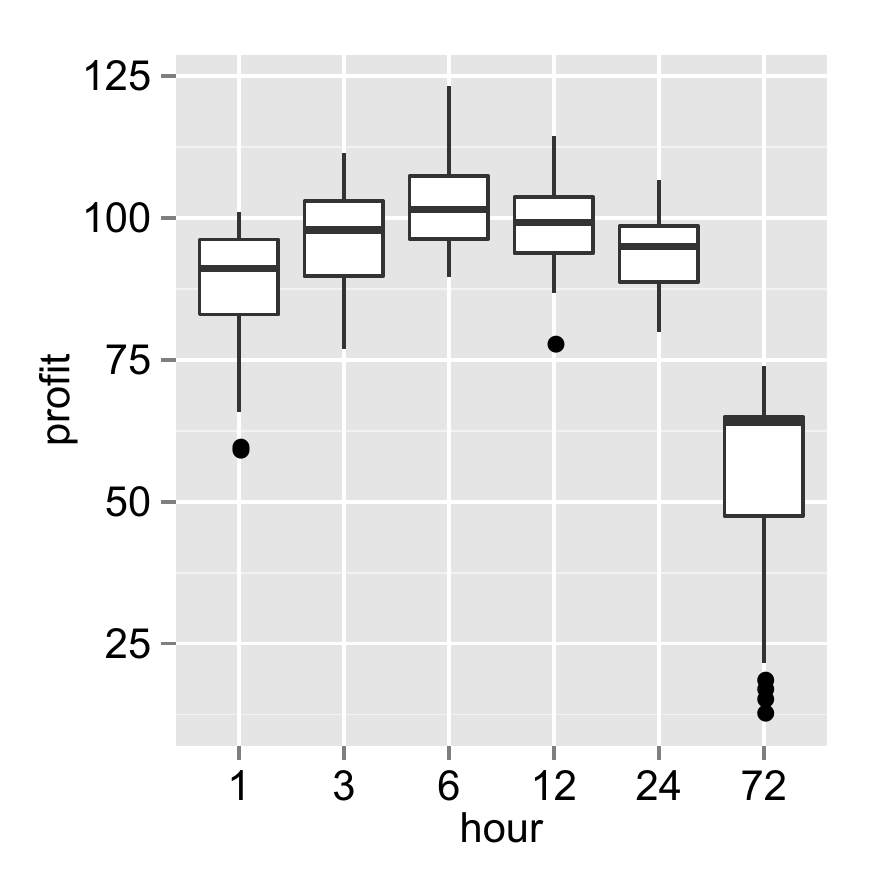}\label{fig:stage-3-profit-hour-hard}}
\vspace{-10pt}
\caption{Dynamic multi-campaign arbitrage net profit distribution with different update frequency.}\label{fig:stage-3-hour}
\end{figure}

In practice, as the market competition and the user behaviour change across the time, the meta-bidder should dynamically change its bidding strategy and campaign selection. In this part of experiment, we test the capability of our proposed \tsf{sam2} bidding strategy with dynamic campaign portfolio selection over a 72 hour test period. The arbitrage bidding function and campaign selection probability are updated periodically, and we refer the interval between two updates as one round. Specifically, at the beginning of each round, we re-train the arbitrage bidding function and campaign selection probability using Algorithm~\ref{algo:sam} based on the bidding data collected from previous round. A problem here is that how frequent the update should be? It is apparent that if the round period is too long, it is difficult for the meta-bidder to catch the transient statistical arbitrage opportunities; if the round period is too short, the training data could be sparse and the model might overfit the data.

We test the dynamic multiple campaign arbitrage on 5 portfolios, each of which consists of 4 campaigns with the data logged within the same period. For each test campaign portfolio, we try the different update frequencies as well as different risk-averse $\alpha$'s. The box plots \cite{mcgill1978variations} of the arbitrage net profit distribution with different update frequencies under two payoff settings are shown in Figure~\ref{fig:stage-3-hour}. From the results we observe that (i) the positive net profit values over all cases demonstrate the capability of \tsf{sam2} to make dynamic arbitrages. (ii) In both payoff settings, the dynamic SAM (period no more than 24 hours) have much better performance than the static SAM (period equals to 72 hours, i.e., only one update), which indicates the importance of dynamically re-training the models to catch the latest market situation. (ii) Among the different frequencies of dynamic updating, updating every 6 hours leads to the highest arbitrage net profit. We believe this is a trade-off point between the abundance and recency of the training data. Note that the optimal update frequency may be different for other campaigns or different training settings.

\begin{figure}[t]
\centering
\vspace{-5pt}
\includegraphics[width=0.9\columnwidth]{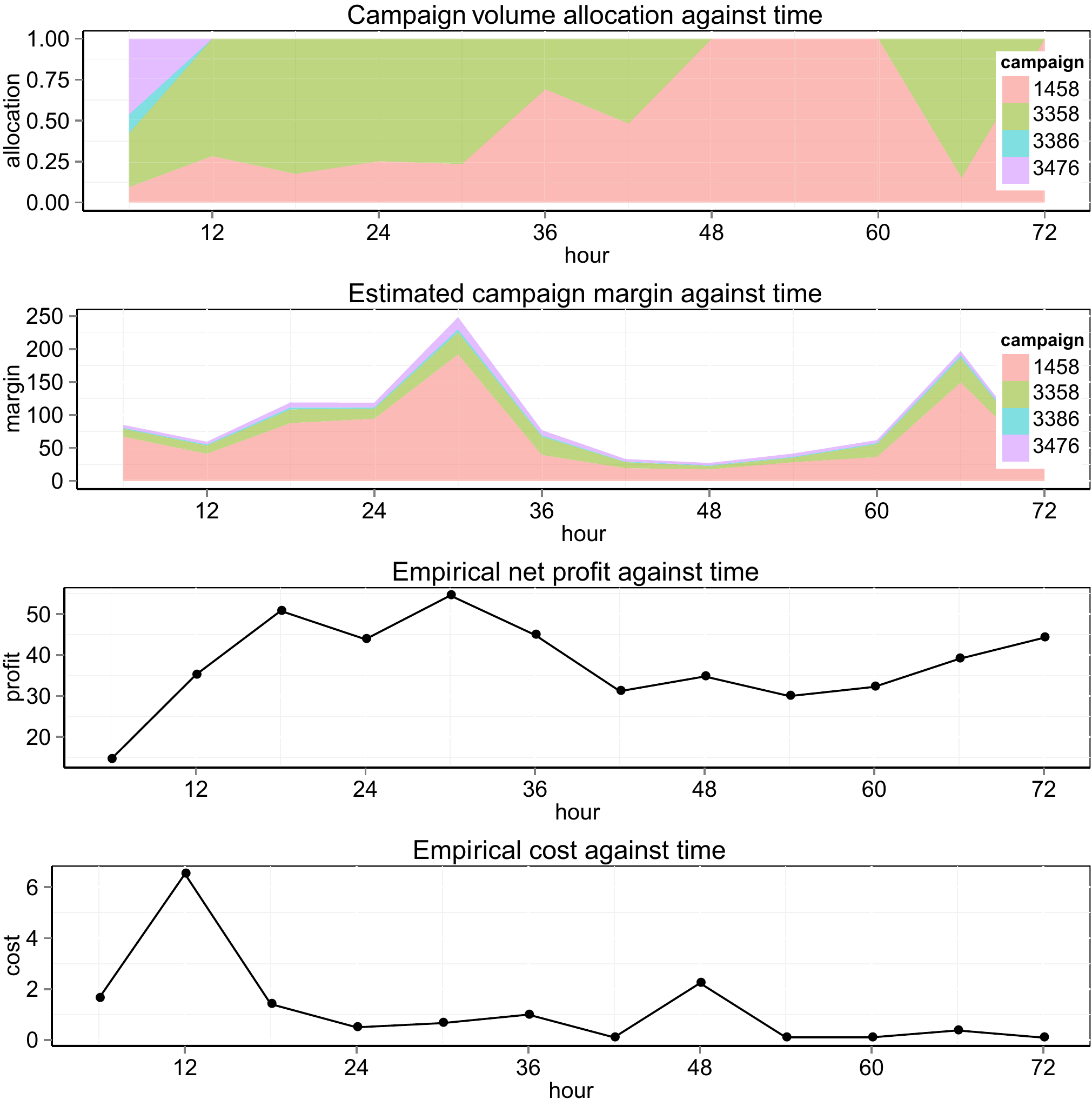}
\vspace{-10pt}
\caption{A case study of dynamic multi-campaign arbitrage performance and the corresponding margin estimation and volume allocation.}\label{fig:stage-3-case}
\vspace{-10pt}
\end{figure}

In addition, Figure~\ref{fig:stage-3-case} presents a case study of the 72 hour dynamic 4-campaign arbitrage with the model updated for every 6 hours. In each round, the calculated campaign selection probability (i.e., the volume allocation) from portfolio optimisation, the estimated net profit margin of each campaign, the empirical net profit and cost are depicted. We observe that the estimated margin for each campaign varies over time, which results in the change of campaign volume allocation across the time. The empirical profit shows the same trend with the estimated campaign margin, which to some extent highlights the effectiveness of the margin estimation in our model. Moreover, the cost in each round (i.e., 6 hours) is different, not necessarily be the average budget allocated for each round. It is possible that if the market is too competitive to make arbitrage profit, the resulting cost and profit could be both much low.

\vspace{-3pt}
\subsection{Online Test}\label{sec:online}
Our SAM algorithm has been deployed and tested in a live environment provided by BigTree DSP. The model training follows the scheme in Section~\ref{sec:multi-cam-arbitrage}. Specifically, with Algorithm~\ref{algo:sam}, we obtain the empirically optimal \tsf{sam2} bidding function $b(\theta, r)$ and campaign selection probability $\bs{v}$ for the meta-bidder based on the 3-campaign training data described in Section~\ref{sec:dataset}, where the hyperparameter $\alpha$ in Eq.~(\ref{eq:opt-obj}) is set as 0.1. As a control baseline, we deployed another meta-bidder with the basic linear bidding function \cite{perlich2012bid,lee2012estimating} and the internal auction-based campaign selection scheme \cite{yuan2013real}, denoted as \textsf{base}. During the online A/B test, every received bid request from the router of BigTree DSP will be randomly assigned to either of the two meta-bidders, which returns the bid response, including the selected campaign ad and the bid price, back to the ad exchange for auction. The online test is conducted during 23 hours between 13 and 14 Feb. 2015 with \$60 budget for each meta-bidder.

\begin{figure}[t]
\centering
\vspace{-6pt}
\includegraphics[width=.9\columnwidth]{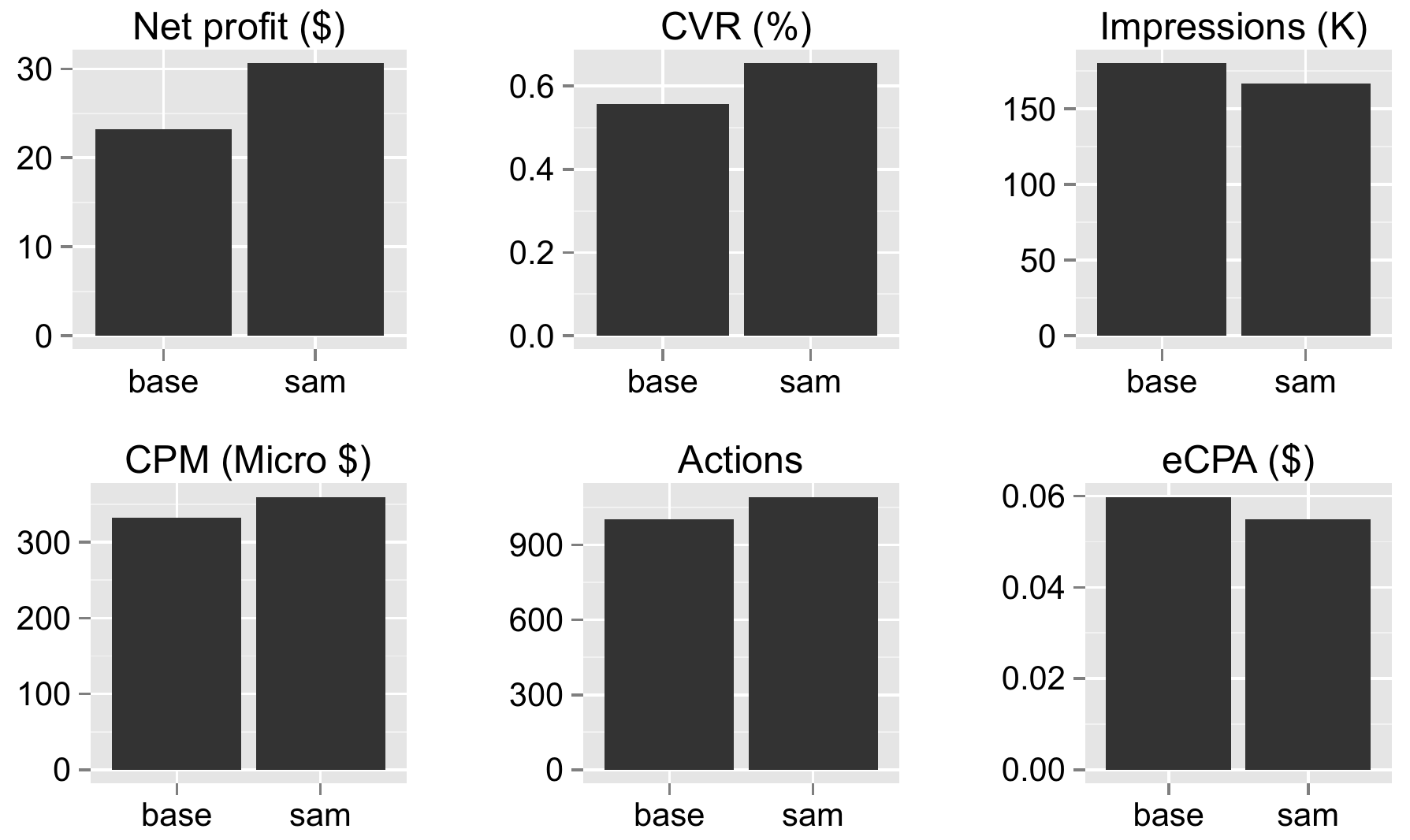}
\vspace{-10pt}
\caption{Online performance on BigTree DSP.}\label{fig:stage-4-online}
\end{figure}

Figure~\ref{fig:stage-4-online} presents the overall online performance of \tsf{sam} and the baseline algorithm \tsf{base}. The online results on the commercial DSP verify the effectiveness of our algorithm in a real commercial setting: \tsf{sam} leads to \$30.6 arbitrage net profit with \$60 budget, which is a 51.1\% margin and a 31.8\% improvement over the \tsf{base} bidder setting. An interesting observation is that in spite of the higher CPM, \tsf{sam} brings lower eCPA than \tsf{base}, which ultimately leads to higher arbitrage net profit. This suggests that despite the market price and arbitrage margin are different across the campaigns, our SAM algorithm would be able to successfully identify and target to the cases that have higher arbitrage margin from those high value impressions (reflected by their high CPM).

\vspace{-5pt}
\section{Conclusions}\label{sec:con}

In this paper, we conducted the first study on statistical arbitrage mining in RTB display advertising. We proposed a joint optimisation framework to maximise the expected arbitrage net profit with budget and risk constraints, which is then solved in an EM fashion. In the E-step the bid volume is reallocated according to the individual campaign's estimated risk and return, while in the M-step the arbitrage bidding function is optimised to maximise the expected arbitrage net profit with the campaign volume allocation. Aside from the theoretical insights, the offline and online large-scale experiments with real-world data demonstrated the effectiveness of our proposed solution in exploiting arbitrage in various model settings and market environments.
We believe this would open up a whole new set of research questions that intersect between financial methods such as high-frequency trading \cite{gatev2006pairs}, risk-management \cite{markowitz1952portfolio,elton2009modern} and data mining methodologies for display advertising and beyond.  In the future work, we plan to further improve the dynamic nature of the SAM model and extend it to mine arbitrage in other domains such as cloud computing and e-commence.

\vspace{5pt}
\noindent
\textbf{Acknowledgement.} We thank the engineers Tianchi Zhu, Jie Liu, Lei Gong, Yanjun He and Zhao Yang for helping conduct the online experiment on BigTree DSP.

\bibliographystyle{abbrv}
\bibliography{fp056-zhang}

\end{document}